\documentclass[reprint,amsmath,amssymb,aps,prd,nofootinbib,floatfix]{revtex4-1}

\usepackage{amssymb,amsmath,amsfonts}
\usepackage{mathrsfs}
\usepackage{dsfont} 
\usepackage{graphicx,bm}
\usepackage{multirow}
\usepackage{color}
\usepackage{enumerate}
\usepackage{placeins}

\usepackage{dcolumn}
\usepackage{array}

\newcommand{\tr}{\mathop{\mathrm{Tr}}}
\newcommand{\nom}{{\nonumber}}
\DeclareMathOperator{\sgn}{sgn}
\providecommand{\e}[1]{\ensuremath{{\scriptscriptstyle E\negthinspace #1}}}

\allowdisplaybreaks

\begin{document}

\title{Neutron star properties with careful parameterization in the (axial)vector meson extended linear sigma model}

\author{P{\'e}ter Kov{\'a}cs}
\email{kovacs.peter@wigner.hu}
\affiliation{
Institute for Particle and Nuclear Physics, Wigner Research Centre for Physics, 1121 Budapest, Hungary
}
\affiliation{
Institute of Physics, E\"otv\"os University, 1117 Budapest, Hungary
}
\author{J{\'a}nos Tak{\'a}tsy}
\affiliation{
Institute for Particle and Nuclear Physics, Wigner Research Centre for Physics, 1121 Budapest, Hungary
}
\affiliation{
Institute of Physics, E\"otv\"os University, 1117 Budapest, Hungary
}
\author{J{\"u}rgen Schaffner-Bielich}
\affiliation{
Institut f{\"u}r Theoretische Physik, Goethe Universit{\"a}t Frankfurt, D-60438 Frankfurt, Germany
}
\author{Gy{\"o}rgy Wolf}
\affiliation{
Institute for Particle and Nuclear Physics, Wigner Research Centre for Physics, 1121 Budapest, Hungary
}

\begin{abstract}
The existence of quark matter inside the cores of heavy neutron stars is a possibility which can be probed with modern astrophysical observations. We use an (axial)vector meson extended quark-meson model to describe quark matter in the core of neutron stars. We discover that an additional parameter constraint is necessary in the quark model to ensure chiral restoration at high densities. By investigating hybrid star sequences with various parameter sets we show that low sigma meson masses are needed to fulfill the upper radius constraints, and that the maximum mass of stable hybrid stars is only slightly dependent on the parameters of the crossover-type phase transition. Using this observation and results from recent astrophysical measurements a constraint of $2.6<g_V<4.3$ is set for the constituent quark -- vector meson coupling. The effect of a nonzero bag constant is also investigated and we observe that its effect is small for values adopted in previous works.
\end{abstract}

\maketitle

\section{Introduction}
\label{sec:intro}

The theory of strong interaction, Quantum Chromodynamics (QCD) is notoriously difficult to tackle, especially at high but not very high densities or temperatures (around the chiral phase boundary). Although lattice Monte-Carlo calculations in the last decades achieved immense progress by solving QCD at low densities and revealing the nature of strongly interacting matter \cite{Aoki2006,Borsanyi2013}, the Sign Problem hinders their application at high densities \cite{deForcrand2010}. On the other hand, perturbative methods only become reliable at very high energies, not relevant for most physical scenarios involving dense nuclear matter \cite{Kurkela2009}. From the experimental side ALICE at CERN \cite{ALICE}, and PHENIX and STAR at RHIC \cite{PHENIX,STAR} also managed to explore QCD at low density and high temperature. Up to this day experimental data at high densities is scarce and have rather bad statistics \cite{STAR,NA61}, however, multiple experimental facilities that are under construction are designed to explore this region with higher precision in the near future \cite{CBM,NICA}.

One possibility to explore this area is the usage of effective theories, which can be applied at finite density and provide important insight about certain aspects of strongly interacting matter. In our approach the underlying principle of constructing such a model is to require that the Lagrangian -- involving composite particles (mesons and constituent quarks) instead of the fundamental quarks and gluons -- has the same global symmetries as QCD itself. One group of these models consists of chiral effective field theories, which are designed to describe chiral restoration at high temperatures and densities, and are expected to be reliable in the vicinity of chiral phase transition.

Recent studies, based on astrophysical measurements in the last few years, argue that deconfined quark matter might also exist inside the core of neutron stars (NSs), and in dense stellar remnants (see \textit{e.g.} \cite{Annala2019,Blaschke2020}). The emergence of increasingly robust predictions were made possible by major advances in the observation of NSs in the previous decade, which have already put multiple constraints on the equation of state (EoS) of dense strongly interacting matter. These constraints stem from a variety of astrophysical observations, ranging from the discovery of NSs with masses of $2$~$M_\odot$ \cite{Demorest2010,Antoniadis2013,Cromartie2019}, through gravitational-wave measurements of the inspirals of NS--NS systems \cite{Abbott2018,Abbott2020}, to the X-ray pulse-profile measurements of pulsars with NICER \cite{Riley2019,Miller2019,Riley2021,Miller2021} together with qualitative improvements in X-ray radius measurements (see \textit{e.g.} \cite{Ozel2015}).

The description of quark matter in connection with neutron stars has been investigated using differing chiral approaches.
The chiral mean field model is based on a Yukawa-type scalar and vector meson exchange for nucleons and quarks obeying chiral symmetry \cite{Dexheimer:2009hi,Negreiros:2010hk,Dexheimer:2012eu,Dexheimer:2018dhb,Roark:2018uls,Dexheimer:2020rlp}.
The chiral quark-meson model uses the linear sigma model and has been extended by implementing vector meson exchange and a vacuum term \cite{Zacchi:2015lwa,Zacchi:2016tjw}. Both additions to the linear sigma model turned out to be of having a significant impact on the properties of compact star configurations.
The quark-meson model has been also investigated for compact stars by going beyond mean-field using renormalization for the quark part \cite{Zacchi:2019ayh}
and by adopting the functional renormalization group method \cite{Otto:2019zjy,Otto:2020hoz}. Confronting recent astrophysical data on neutron stars
with chiral models for quark matter has been also performed within a nonlocal Nambu-Jona-Lasinio model \cite{Alvarez-Castillo:2018pve,Shahrbaf:2019vtf,Blaschke:2020qqj}
and a unified quark-meson-nucleon model \cite{Marczenko:2020jma}.

In this paper we use an (axial)vector meson extended linear sigma model with constituent quarks (or quark-meson model) at zero temperature and finite quark (or baryon) chemical potential to describe the properties of hybrid stars. The advantage of this model -- altogether with the parameterization procedure and the approximations that were used -- is that it reproduces the meson spectrum (and also various decay widths) quite well at $T=\mu_q=0$ \cite{Parganlija:2012fy} and moreover its finite temperature version also agrees well with various lattice results \cite{Kovacs:2016juc}. Since we think that parameterization plays a crucial role -- i.e. depending on the starting position of the parameter space the system shows very different behavior at finite $T$ and/or $\mu_q$ -- in the description of properties at finite densities, we investigate the consequences of the asymptotic behavior of the system of equations on the parameterization. It turns out that the system does not behave as expected with every parameterization. 

Using two different hadronic models at low densities we construct hybrid star EoS's, which fulfill all the current expectations coming from astrophysical measurements, providing some of the parameters, like the $g_V$ vector coupling, are set properly.

The paper is organized as follows. Section \ref{sec:eLSM} is devoted to the introduction of the model, set up of the $\beta$-equilibrium and charge neutrality conditions, calculation of the pressure and field equations, and the parameterization procedure. In Section \ref{sec:compstar} the hadronic EoS's and interpolation methods (between the hadronic and the quark EoS's) are demonstrated together with a brief summary on compact star observables. Section \ref{sec:results} contains our results, where the EoS's, the $M-R$ curves, the tidal deformabilities ($\Lambda$) and their dependence on various parameters are analyzed. Finally, we summarize the implications of our work in Section \ref{sec:conclusion}. Some additional details can be found in Appendices \ref{app:omega}-\ref{app:param}.

\section{The vector meson extended linear sigma model}
\label{sec:eLSM}

The Lagrangian of the model is a version of the three flavored (axial)vector meson extended linear sigma model introduced in \cite{Parganlija:2012fy}, where it was thoroughly investigated at zero temperature. A slightly modified version of that model was used for finite temperature investigations in \cite{Kovacs:2016juc}. Here we use the latter with an additional (axial)vector Yukawa type term.  Consequently, the total Lagrangian of the model reads as
\begin{widetext}
\begin{align}
  \mathcal{L} & = \tr[(D_{\mu}M)^{\dagger}(D_{\mu}M)] -
  m_{0}^{2}\tr(M^{\dagger}M) - \lambda_{1}[\tr(M^{\dagger} M)]^{2} - \lambda_{2}\tr(M^{\dagger}M)^{2} 
  + c_{1}(\det M+\det M^{\dagger}) + \tr[H(M+M^{\dagger})] \nom \\
  & -\frac{1}{4}\tr(L_{\mu\nu}^{2}+R_{\mu\nu}^{2}) 
  + \tr\left[ \left(\frac{m_{1}^{2}}{2}+\Delta\right) (L_{\mu}^{2}+R_{\mu}^{2})\right] 
  + i\frac{g_{2}}{2}(\tr\{L_{\mu\nu}[L^{\mu},L^{\nu}]\} + \tr\{R_{\mu\nu}[R^{\mu},R^{\nu}]\})\nom \\
  & +\frac{h_{1}}{2}\tr(M^{\dagger}M)\tr(L_{\mu} ^{2}+R_{\mu}^{2})
  + h_{2}\tr(|L_{\mu}M|^{2}+|M R_{\mu} |^{2}) + 2h_{3}\tr(L_{\mu}M R^{\mu}M^{\dagger}) 
    + \bar{\Psi}\left[i \gamma_{\mu}D^{\mu}-\mathcal{M}\right]\Psi \nom \\
  & - g_V \bar{\Psi}\left[\gamma_{\mu}V^{\mu} + \frac{g_A}{g_V}\gamma^{5} \gamma_{\mu}A^{\mu}\right]\Psi\, , 
\label{Eq:Lagr}
\end{align}
\end{widetext}
where, as it is described in detail in \cite{Kovacs:2016juc}, $M=M_S + M_{PS}$, $L^{\mu}=V^{\mu}+A^{\mu}$,  $R^{\mu}=V^{\mu}-A^{\mu}$ and $M_S, M_{PS}, V^{\mu}, A^{\mu}$ stand for the scalar, the pseudoscalar,  the vector and the axial vector nonets, respectively, while $\Psi = (q_u, q_d, q_s)^T$ for the constituent quark fields. Some comments are in order: there are two new unknown parameters, the $g_V$ vector coupling, which has a direct impact on the value of the maximal mass on the $M-R$ curve of compact stars. Although the $g_A$ axial coupling (or the $g_A/g_V$ ratio) is also unknown it will not appear in any of the following expressions, thus its value is irrelevant for the current investigation. In \cite{Kovacs:2016juc} we also introduced Polyakov-loop variables, which vanish at zero temperature, thus they will not be present in the equations relevant for compact stars, however, since in the parameterization the pseudocritical temperature is used, it still affects the parameter set.

As a standard procedure for theories with spontaneous symmetry breaking (SSB), nonzero vacuum expectation values (vev) are assumed for the non-strange and strange isoscalar fields $\sigma_N$ and $\sigma_S$ and for the temporal component of the three vector fields: the charge neutral $\rho_0^{\mu}$, $\omega^{\mu}$, and $\Phi^{\mu}$. These expectation values are denoted as follows:
\begin{align}
 \langle \sigma_N\rangle &\equiv \phi_N,\quad \langle \sigma_S\rangle \equiv \phi_S, \nom\\ \langle \rho_0^{0}\rangle &\equiv v_{\rho},\quad \langle \omega^{0}\rangle \equiv v_{\omega},\quad \langle \Phi^{0}\rangle \equiv v_{\Phi}. 
 \label{Eq:vev}
\end{align}
It should be noted that we neglect here the small effect of isospin breaking, which would require the introduction of a nonzero expectation value for the scalar $a_0^0$ field. 

Hereafter the fields are shifted with their nonzero expectation values, which subsequently results in the tree-level expressions for the meson and constituent quark masses and the tree-level decay widths. Moreover, the nonzero vector vev's shift the $\mu_u$, $\mu_d$, and $\mu_s$ quark chemical potentials of the constituent quark fields subsequently leading to the following effective quark chemical potentials for the different flavors,
\begin{align}
    \Tilde{\mu}_u &= \mu_u - \frac{1}{2}g_V(v_{\omega} + v_{\rho}), \nom \\
    \Tilde{\mu}_d &= \mu_d - \frac{1}{2}g_V(v_{\omega} - v_{\rho}),  \\
    \Tilde{\mu}_s &= \mu_s - \frac{1}{\sqrt{2}}g_V v_{\Phi}. \nom
\end{align}
These shifts stem from the vector Yukawa term, that is the last term of the Lagrangian (Eq.~\eqref{Eq:Lagr}).

\subsection{$\beta$-equilibrium and charge neutrality}

We add a free electron gas to our system with some $\mu_e$ electron chemical potential and assume $\beta$-equilibrium, that is 
\begin{equation}
    \mu_d = \mu_s = \mu_u + \mu_e \nom
\end{equation}
after neutrinos have left the system. Thus using 
\begin{equation}
    \mu_q\equiv \frac{1}{3}\mu_B = \frac{1}{3}(\mu_u + \mu_d + \mu_s)
\end{equation}
 for the quark chemical potential, the chemical potentials for the different flavors are given by
\begin{align}
    \mu_u &= \mu_q - \frac{2}{3}\mu_e, \nom \\
    \mu_d &= \mu_q + \frac{1}{3}\mu_e,\label{Eq:mu_uds} \\
    \mu_s &= \mu_q + \frac{1}{3}\mu_e. \nom
\end{align}

Charge neutrality is also applied, which can be written as
\begin{equation}
    \frac{2}{3}n_u-\frac{1}{3}n_d-\frac{1}{3}n_s-n_e = 0,
\end{equation}
where $n_{u/d/s}$ and $n_e$ are the number densities for the $u,d,s$ quarks and the electron, respectively. They can be calculated as
\begin{equation}
    n_f = \frac{\partial p}{\partial \mu_f},\, f\in(u,d,s),\quad n_e = \frac{\partial p}{\partial \mu_e},
\end{equation}
where $p$ is the pressure. 

Finally, the effective quark chemical potentials for the different flavors can be written as,
\begin{align}
    \Tilde{\mu}_u &= \mu_q - \frac{2}{3}\mu_e - \frac{1}{2}g_V(v_{\omega} + v_{\rho}), \nom \\
    \Tilde{\mu}_d &= \mu_q + \frac{1}{3}\mu_e - \frac{1}{2}g_V(v_{\omega} - v_{\rho}),\label{Eq:mu_eff_uds} \\
    \Tilde{\mu}_s &= \mu_q + \frac{1}{3}\mu_e - \frac{1}{\sqrt{2}}g_V v_{\Phi}. \nom
\end{align}
It should be noted here that while $\mu_{u/d/s}$ play the role of the usual chemical potentials, in the calculation of the grand potential (or the pressure) the $\Tilde{\mu}_{u/d/s}$ effective chemical potentials appear.

\subsection{Pressure and the field equations}

The pressure is given by
\begin{equation}
    p(\mu_f, \mu_e) = \Omega_0 - \Omega(T=0, \mu_f, \mu_e),
    \label{Eq:pressure}
\end{equation}
where $\Omega$ is the grand potential. The grand potential is calculated in a hybrid approximation used in \cite{Kovacs:2016juc} at zero temperature with additional vector condensates introduced above. In this hybrid approach we consider only quark fluctuations, while all the mesons are at tree-level. The grand potential consist of the following terms,
\begin{equation}
    \Omega_{\textrm{tot}} = U_{\textrm{meson}}(\phi_N, \phi_S, v_{\rho}, v_{\omega}, v_{\Phi}) + \Omega_{\bar q q}^{\textrm{vac}} + \Omega_{\bar q q}^{\textrm{mat}}(\mu_f) + \Omega_{\textrm{el}},
\end{equation}
where $U_{\textrm{meson}}$ stands for the tree\,-\,level meson potential,  $\Omega_{\bar q q}^{\textrm{vac}}$ and $\Omega_{\bar q q}^{\textrm{mat}}$ for the vacuum and matter part of the one\,-\,loop constituent quark contributions, while $\Omega_{\textrm{el}}$ for the electron contribution. Its explicit form can be found in Eq.~\eqref{Eq:GP} of Appendix~\ref{app:omega}. The field equations (FE) are the stationary points of the grand potential, {\it i.e.}
\begin{equation}
    \frac{\partial\Omega_{\textrm{tot}}}{\partial \phi_N} = \frac{\partial\Omega_{\textrm{tot}}}{\partial \phi_S} = \frac{\partial\Omega_{\textrm{tot}}}{\partial v_{\omega}} = 
    \frac{\partial\Omega_{\textrm{tot}}}{\partial v_{\rho}} = 
    \frac{\partial\Omega_{\textrm{tot}}}{\partial v_{\Phi}} = 0.
\end{equation}
These are five coupled equations for the $\mu_q$ (or $\mu_B$) dependence of the scalar and vector condensates. There is another unknown, the $\mu_e$ electron chemical potential, which is determined through the charge neutrality condition, which is also coupled to the preceding five equations. The explicit form of the system of six equations that needs to be solved reads as
\begin{widetext}
  \begin{align}
    \frac{\partial\Omega_{\textrm{tot}}}{\partial \phi_N} &= m_0^2\phi_N + \lambda_1 \left(\phi_N^2 + \phi_S^2 \right)\phi_N + \frac{1}{2}\lambda_2 \phi_N^3 - \frac{c_1}{\sqrt{2}} \phi_N \phi_S -h_N - \frac{1}{2}\left(h_1 + h_2 + h_3\right) \phi_N\left(v_{\omega}^2 + v_{\rho}^2\right) - \frac{1}{2}h_1 \phi_N v_{\Phi}^2 \nom\\
    -&\frac{3g_F}{8\pi^2}m_u^3\left[1 + 4 \log\frac{m_u}{M_0} \right] + \frac{3g_F}{4\pi^2} \sum_{f\in (u,d)} m_f^3\left[ \gamma_f \sqrt{\gamma_f^2 - 1} - \log\left( \gamma_f + \sqrt{\gamma_f^2 - 1} \right) \right] = 0 \label{Eq:FE_N}\\
    \frac{\partial\Omega_{\textrm{tot}}}{\partial \phi_S} &= m_0^2\phi_S + \lambda_1 \left(\phi_N^2 + \phi_S^2 \right)\phi_S + \lambda_2 \phi_S^3 - \frac{c_1}{2\sqrt{2}} \phi_N^2 -h_S - \frac{1}{2}h_1 \phi_S\left(v_{\omega}^2 + v_{\rho}^2\right) - \frac{1}{2}\left(h_1 + 2h_2 + 2 h_3\right) \phi_S v_{\Phi}^2 \nom \\
    -&\frac{3g_F}{8\sqrt{2}\pi^2}m_s^3\left[1 + 4 \log\frac{m_u}{M_0} \right] + \frac{3g_F}{2\sqrt{2}\pi^2} m_s^3\left[ \gamma_s \sqrt{\gamma_s^2 - 1} - \log\left( \gamma_s + \sqrt{\gamma_s^2 - 1} \right) \right] = 0 \label{Eq:FE_S}\\ 
    \frac{\partial\Omega_{\textrm{tot}}}{\partial v_{\omega}} &= -m_\rho^2 v_{\omega} + \frac{m_u^3}{2\pi^2}g_V \sum_{f \in (u,d)}\sgn(\Tilde{\mu}_f) \left(\gamma_f^2 - 1\right)^{\frac{3}{2}} =0 \label{Eq:FE_om} \\ 
    \frac{\partial\Omega_{\textrm{tot}}}{\partial v_{\rho}} &= -m_\rho^2 v_{\rho} + \frac{m_u^3}{2\pi^2}g_V \left[ \sgn(\Tilde{\mu}_u) \left(\gamma_u^2 - 1\right)^{\frac{3}{2}} - \sgn(\Tilde{\mu}_d) \left(\gamma_d^2 - 1\right)^{\frac{3}{2}}\right] =0 \label{Eq:FE_rh}\\ 
    \frac{\partial\Omega_{\textrm{tot}}}{\partial v_{\Phi}} &= -m_\Phi^2 v_{\Phi} + \frac{m_s^3}{\sqrt{2}\pi^2}g_V \sgn(\Tilde{\mu}_s) \left(\gamma_s^2 - 1\right)^{\frac{3}{2}} = 0 \label{Eq:FE_ph} \\
    2 \sgn&(\Tilde{\mu}_u)\left(\Tilde{\mu}_u^2 - m_u^2\right)^{\frac{3}{2}} - \sgn(\Tilde{\mu}_d)\left(\Tilde{\mu}_d^2 - m_d^2\right)^{\frac{3}{2}} - \sgn(\Tilde{\mu}_s)\left(\Tilde{\mu}_s^2 - m_s^2\right)^{\frac{3}{2}} - \sgn(\mu_e)\left(\mu_e^2 - m_e^2\right)^{\frac{3}{2}} = 0,
    \label{Eq:FE_ch}
  \end{align}
\end{widetext}
where $\gamma_f$ is defined in Eq.~\eqref{Eq:gamma_f}, while the $m_\rho=m_\omega$, and $m_\Phi$ masses are given in Eqs.~\eqref{Eq:mrho}, \eqref{Eq:mphi}. It should be noted that each term that contains $\sqrt{\gamma_f^2-1}$ or $\sqrt{\Tilde{\mu}_f^2-m_f^2}$ is only present if $\Tilde{\mu}_f > m_f$. 

\subsection{Model parameters and asymptotic behavior of the equations}

In order to solve our system of equations (Eqs.~\eqref{Eq:FE_N}-\eqref{Eq:FE_ch}) at finite $\mu_q$, the parameters of the model should be determined first. The 15 unknown parameters are $m_0$, $\lambda_1$, $\lambda_2$, $c_1$, $m_1$, $g_1$, $g_2$, $h_1$, $h_2$, $h_3$, $\delta_S$, $\phi_N$, $\phi_S$, $g_F$, and $g_V$. As described in detail in Section~IV. of \cite{Kovacs:2016juc} we calculate meson masses and decay widths at $\mu_q=T=0$, moreover $T_c$ at $\mu_q=0$ and compare them to their experimental value -- taken from the PDG \cite{Zyla:2020zbs}, while for $T_c$ the value $150$~MeV was used, which was taken from the lattice \cite{AOKI200646, Borsanyi:2010bp} --  through a $\chi^2$ fit. Since in the current approximation we use tree\,-\,level (axial)vector masses the $g_V$ vector coupling does not appear in any of the expressions, thus it remains a free parameter. It should be noted, however, that if we include fermionic fluctuations in the curvature masses of the (axial)vectors, $g_V$ will also be fixed through the $\chi^2$ minimization \cite{Kovacs:2021kas}.

\begin{figure}[htbp]
  \centering
  \includegraphics[width=0.48\textwidth]{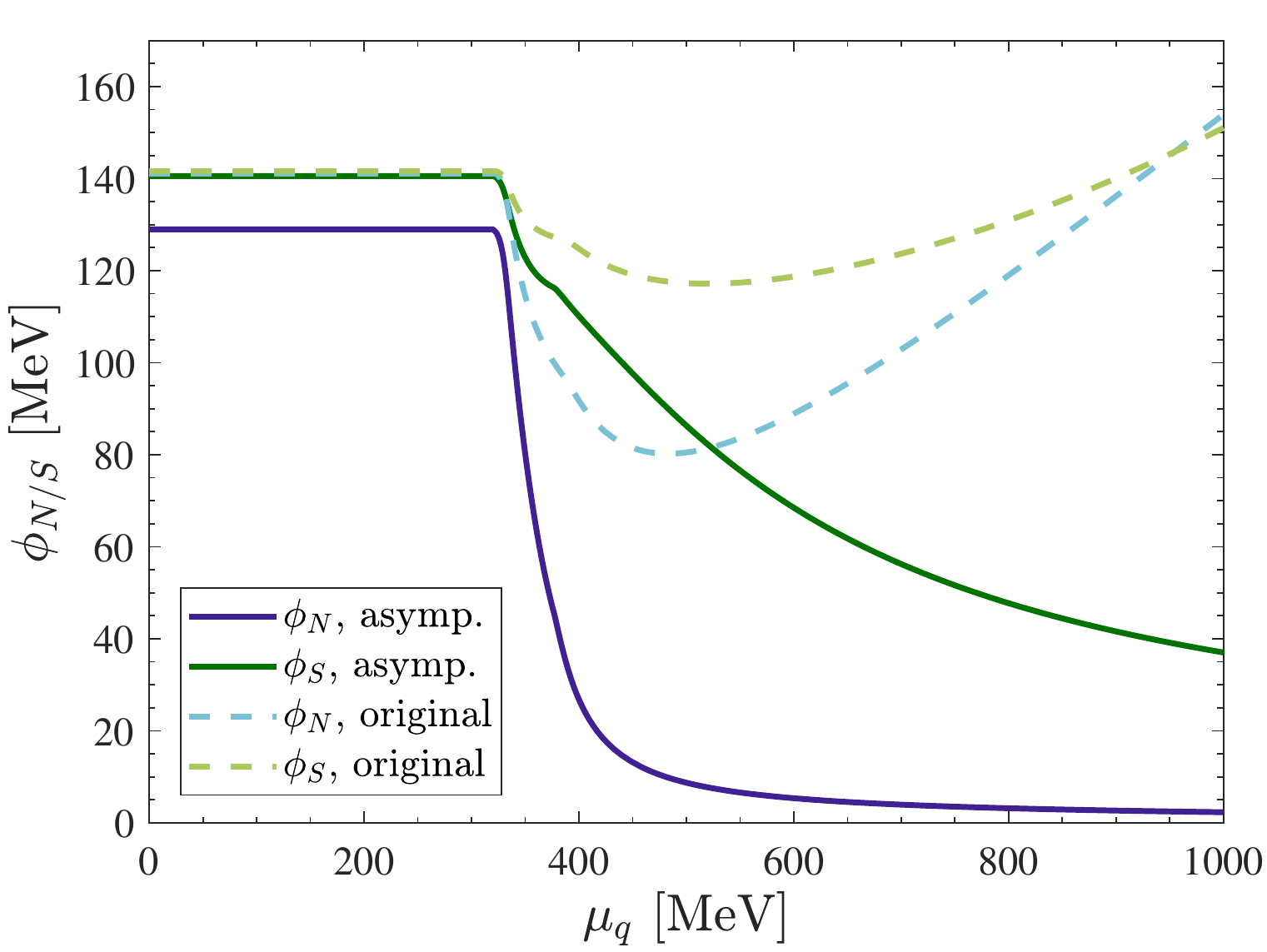}
  \caption{\label{fig:phiNS}The dependence of the $\phi_{N/S}$ scalar condensates on the quark chemical potential with (solid lines) and without (dashed lines) the asymptotic condition of Eq.~\eqref{Eq;asym_cond}. The parameter sets for these results are shown in Table~\ref{Tab:param1}, and in Table~IV. of \cite{Kovacs:2016juc}, respectively, and $g_V=5$ in both cases. This shows that a naive parameterization of the model results in an increase of the scalar condensates at higher densities and therefore the recurrence of chiral symmetry breaking.}
\end{figure}

During the investigation of different parameter sets we noticed some unusual behaviour of the $\phi_{N/S}$ scalar condensates as a function of $\mu_q$, namely for large $\mu_q$ values $\phi_{N/S}$ started to increase substantially, as shown in Fig.~\ref{fig:phiNS}. This behavior would mean the recurrence of chiral symmetry breaking for large $\mu_q$, which seems unphysical. To avoid this unwanted behavior we investigated the asymptotic solution for large $\mu_q$ and demanded the disappearance of the $\phi_{N/S}$ condensates at these asymptotic values. Thus, more explicitly it is assumed that for large $\mu_q$
\begin{align}
    &\phi_N\sim \mu_q^{-\alpha},\;\phi_S\sim \mu_q^{-\beta},\nom\\ 
    &v_{\omega}\sim \mu_q^{k},\; v_{\rho}\sim \mu_q^{l},\; v_{\Phi}\sim \mu_q^{n},\nom\\
    &\textrm{with}\; \alpha, \beta, k, l, n \geq 0.\label{Eq:cond_mu_q_scaling}
\end{align}
It is also assumed that $\Tilde{\mu}_{u/d/s} > 0$. With these assumptions in leading order
\begin{align*}
    \sgn(\Tilde{\mu}_f) m_f^3(\gamma_u^2 - 1)^{\frac{3}{2}}&\approx \Tilde{\mu}_f^3,\\
    m_{\omega/\rho/\Phi}^2 &\approx m_1^2.
\end{align*}
Consequently, Eqs.~\eqref{Eq:FE_om}-\eqref{Eq:FE_ch} will have the forms -- neglecting the electron's mass as well,
\begin{align}
    m_1^2v_{\omega} &\approx \frac{g_V}{2\pi^2}\left(\Tilde{\mu}_u^3 + \Tilde{\mu}_d^3\right),\\
    m_1^2v_{\rho} &\approx \frac{g_V}{2\pi^2}\left(\Tilde{\mu}_u^3 - \Tilde{\mu}_d^3\right),\\
    m_1^2v_{\Phi} &\approx \frac{g_V}{\sqrt{2}\pi^2}\Tilde{\mu}_s^3,\label{Eq:approx_cond_Phi}\\
    \mu_e^3 &\approx 2\Tilde{\mu}_u^3 - \Tilde{\mu}_d^3 - \Tilde{\mu}_s^3. \label{Eq:approx_cond_mue}
\end{align}
Rearranging Eq.~\eqref{Eq:approx_cond_Phi} and substituting Eqs.~\eqref{Eq:mu_eff_uds} for $\Tilde{\mu}_s$ it can be written that
\begin{equation}
       \frac{g_V}{\sqrt{2}\pi^2}\left(\mu_q + \frac{1}{3}\mu_e - \frac{1}{\sqrt{2}}g_V v_{\Phi} \right)^3 - m_1^2v_{\Phi} \approx 0,
\end{equation}
where the first part of the left hand side is a third order polynomial in $v_{\Phi}$ that scales with $\sim \mu_q^{3n}$, thus the second term $m_1^2v_{\Phi}$ which scales only with $\sim \mu_q^{n}$ can be neglected compared to that since $n\geq 0$. Consequently, the expression in the parentheses -- which is simply $\Tilde{\mu}_s$ -- is zero in leading order ($\Tilde{\mu}_s \approx 0$). Similar arguments lead to $\Tilde{\mu}_u \approx 0,\; \Tilde{\mu}_d \approx 0$. Accordingly, from Eq.~\eqref{Eq:approx_cond_mue} we conclude that $\mu_e \approx 0$. Thus the effective quark chemical potentials for the flavors in leading order are zero, 
\begin{equation}
    \Tilde{\mu}_u \approx 0,\; \Tilde{\mu}_d \approx 0,\; \Tilde{\mu}_s \approx 0\; \mu_e \approx 0.
    \label{Eq:mu_U_d_s_cond}
\end{equation}
Using Eq.~\eqref{Eq:mu_eff_uds} in Eqs.~\eqref{Eq:mu_U_d_s_cond} and rearranging for the vector condensates results in
\begin{align}
    v_{\omega} &\approx \frac{2}{g_V}\mu_q\\
    v_{\rho} &\approx 0\\
    v_{\Phi} &\approx \frac{\sqrt{2}}{g_V}\mu_q.
\end{align}
Using these approximations in Eq.~\eqref{Eq:FE_N} gives
\begin{equation}
    -h_N - \frac{1}{2}(h_1 + h_2 + h_3)\phi_N\frac{4}{g_V^2}\mu_q^2-\frac{1}{2}h_1\phi_N\frac{2}{g_V^2}\mu_q^2 \approx 0.
\end{equation}
Expressing $\phi_N$ finally results in
\begin{equation}
    \phi_N = -\frac{h_N g_V^2}{3h_1 + 2 h_2 + 2 h_3}\mu_q^{-2},
\end{equation}
Since $\phi_N>0$, and $h_N>0$, this implies the following condition:
\begin{equation}
    3h_1 + 2 h_2 + 2 h_3 < 0.
    \label{Eq;asym_cond}
\end{equation}
Investigating the asymptotic behavior of Eq.~\eqref{Eq:FE_S} for $\phi_S$ leads to the same condition. Consequently, the asymptotic exponents of the condensates are:
\begin{equation}
    \alpha = \beta = 2,\quad k=n=1,\quad l=0.
\end{equation}
Note: we also checked this numerically with different parameterizations and found that the condition in \eqref{Eq;asym_cond} is valid.

It is worth to note that to reach that conclusion we assumed in Eq.~\eqref{Eq:cond_mu_q_scaling} that $k,l$ and $n$ exponents are all non-negative. Let us now assume for instance that contrary to that $n<0$. In this case in $\Tilde{\mu}_s$ (Eq.~\eqref{Eq:mu_eff_uds}) the $v_{\Phi}$ term could be neglected compared to $\mu_q$ and we would end up with $m_1^2v_{\Phi} \approx \frac{g_V}{\sqrt{2}\pi^2}(\mu_q + 1/3\mu_e)^3$, where the left hand side tends to zero, while the right hand side tends to infinity. Similarly, the assumptions $k<0$ or $l<0$ would also result in contradictions.

We should add that although with this condition the scalar condensates tend to zero at asymptotically large densities, they can still show an increase at intermediately large densities where the asymptotic behaviour does not yet apply. However, this only happens at large densities, which are not present inside neutron stars, and where our model would already lose its predictive power.  

Returning to the determination of the parameters, we use the same procedure as described in \cite{Kovacs:2016juc} with the additional condition of Eq.~\eqref{Eq;asym_cond}. The resulting parameter set is given in Table~\ref{Tab:param1}.
\begin{table}[!htbp]
\caption{Parameter values for $m_{\sigma} =290$~MeV \label{Tab:param1}}
\centering
\begin{tabular}[c]{|c|c||c|c|}\hline
 Parameter & Value & Parameter & Value \\\hline\hline
 $\phi_{N}$ [GeV]& $0.1290$ & $g_{1}$ & $5.3296$\\\hline
 $\phi_{S}$ [GeV]& $0.1406$ & $g_{2}$ & $-1.0579$\\\hline
 $m_{0}^2$ [GeV$^2$] & $-1.2370\e{-2}$ & $h_{1}$ & $5.8467$\\\hline
 $m_{1}^2$ [GeV$^2$] & $0.5600$ & $h_{2}$ & $-12.3456$\\\hline
 $\lambda_{1}$ & $-1.0096$ & $h_{3}$ & $3.5755$\\\hline
 $\lambda_{2}$ & $25.7328$ & $g_{F}$ & $4.9571$\\\hline
 $c_{1}$ [GeV]& $1.4700$ & $M_{0}$ [GeV]& $0.3935$\\\hline
 $\delta_{S}$ [GeV$^2$] & $0.2305$ & \multicolumn{1}{c}{} & \multicolumn{1}{c}{} \\\cline{1-2}
\end{tabular}
\end{table}

In this case the $\sigma$ (or $f_0$) mass is quite low, namely $m_{\sigma}=290$~MeV. We may also study the dependence of our results on the $\sigma$ mass, since our fit prefers a rather small value for that, and experimentally it is not very well defined either. Indeed it is a very broad resonance: $m_{f_0(500)}=400\,\textrm{to}\, 800$~ MeV, $\Gamma_{f_0(500)}=100\,\textrm{to}\, 800$~MeV \cite{Zyla:2020zbs}. For any chosen $\sigma$ mass we have to find a parameter set,
which reproduces that value. To achieve this we have increased the contribution of the $\sigma$ mass to the $\chi^2$ by a factor of 1000. This way, by minimizing the $\chi^2$, the obtained fit reproduces the prescribed $\sigma$ mass with less than $0.2\%$ error. The resulting parameter sets can be found in Appendix~\ref{app:param}.

\section{Compact stars}
\label{sec:compstar}

\subsection{Hadronic equation of state}

To be able to construct the sequence of stable NSs we need a reliable equation of state covering many orders of magnitudes in density from subsaturation densities up to about $n \approx 5-6 \, n_0$, with $n_0$ being the nuclear saturation density. Below saturation nuclear methods such as hadron resonance gas models \cite{Huovinen2009,Bazavov2012} and chiral effective field theories \cite{Lynn2015,Tews2018} offer a robust way for describing nuclear matter. The uncertainties of chiral effective field theories mainly stem from the truncation of the nuclear Hamiltonian within the expansion, as well as the regulariztaion scheme and scale, resulting in the margins of error for the pressure in state-of-the-art models increasing rapidly above $n\approx 2 \, n_0$ \cite{Tews2019}. Looking at the other side of the density spectrum, at extremely high densities we expect QCD to become asymptotically free, thus enabling the use of perturbative QCD (pQCD) calculations \cite{Mogliacci2013}. However, these methods are only reliable at densities much higher than the ones expected to be present in the center of the most massive NSs. Nevertheless, the EoS of quark matter models should converge to those obtained from pQCD calculations at asymptotically high densities.

In the intermediate-density region therefore no fully reliable theory exists, and indeed there is a large selection of nuclear theories using various approaches that range from variational methods to relativistic mean field (RMF) theories \cite[cite review paper][]{}, which results in a high variation in the calculated nuclear EoS's. These approaches are all based on extrapolations from experimentally measurable nuclear properties at saturation. The most important of these quantities are the binding energy per nucleon ($E_0\approx-16.3$~MeV), compressibility ($K_0 = 240 \pm 20$~MeV), symmetry energy ($S_0 = 31.6 \pm 2.7$~MeV) and the slope of symmetry energy ($L = 58.9 \pm 16$~MeV) \cite{Shlomo2006,Piekarewicz2009,Chen2010,Li2019}.

We use two RMF models, the Steiner--Fischer--Hempel (SFHo) model \cite{Steiner2013} and the density-dependent RMF model of Typel et al. (DD2) \cite{Typel2010,Hempel2009}. They are both consistent with the aforementioned nuclear constraints, the only major difference being the different values for the slope of the symmetry energy $L$. This results in the DD2 EoS being stiffer than the SFHo EoS. Some basic properties of the two models are included in Table~\ref{tab:HEoS}. We assume NSs have hadronic crusts described by the EoS's of Baym et al. \cite{Baym1971} and Negele~\&~Vautherin \cite{Negele1971}.

\begin{table}[!htbp]
  \caption{\label{tab:HEoS}%
Nuclear properties of symmetric nuclear matter described by the SFHo and DD2 RMF models as well as some properties of neutron stars described by these models.
}
  \begin{ruledtabular}
  \begin{tabular}{l|p{0.055\textwidth}|p{0.055\textwidth}}
  \multicolumn{1}{c|}{\textrm{Property}} & \textrm{SFHo} & \textrm{DD2}\\
  \colrule
  Saturation density, $n_0$ [fm$^{-3}$] & 0.16 & 0.15 \\
  Binding energy per baryon, $E_0$ [MeV] & -16.17 & -16.02 \\
  Compressibility, $K_0$ [MeV] & 245.2 & 242.7 \\
  Symmetry energy, $S_0$ [MeV] & 31.2 & 32.73 \\
  Slope of symmetry energy, $L$ [MeV] & 45.7 & 57.94 \\
  Maximum mass neutron star [$M_\odot$] & 2.06 & 2.42 \\
  Radius of $M=1.4\,M_\odot$ neutron star [km] & 11.97 & 13.26 \\
  \end{tabular}
  \end{ruledtabular}
\end{table}

\subsection{Hadron--quark phase transition}
\label{ssec:hq_trans}

Since the hadronic and quark phases are described by qualitatively different models, we need to match them 'by hand' and find an appropriate interpolation method to fix the EoS at intermediate densities. One method uses a simple Maxwell construction, assuming that the hadronic and quark models describe strongly interacting matter correctly below/above the phase transition point, where $p_H(\mu_B)=p_Q(\mu_B)$. This construction results in a first-order phase transition and is limited to cases where the hadronic EoS is stiffer than the quark EoS.

Instead of the Maxwell construction here we use two methods that interpolate between the hadronic and quark EoS's on a finite density range, and hence result in crossover phase transitions. The idea of a smooth interpolation can be supported by the argument that both models lose their validity in the intermediate density region, and therefore a strict extrapolation of the two EoS's is generally not justified.

One of these methods interpolates between the pressures, $p(\mu_B)$, on a finite range of chemical potential \citep[see e.g.][]{Abgaryan2018,Baym2019}. Here the hadronic EoS is restricted to the domain below $\mu_{BL}$, and the quark EoS to the domain above $\mu_{BU}$. These chemical potentials correspond to baryon number densities $n_{L}$ and $n_{U}$, respectively. In the intermediate region a reasonable choice for the interpolating function is a polynomial that smoothly connects the two parts:
\begin{equation}
    p(\mu_B) = \sum_{m=0}^{N} C_m \mu_B^m \:, \quad \mu_{BL} < \mu_B < \mu_{BU} ,
\end{equation}
where $C_m$ are coefficients that we may fix by matching the pressure and its derivatives at the boundary points. We use a fifth-order polynomial, and match the pressure, the number density, and the sound speed at both boundary points (this is equivalent to matching the pressure together with its first and second derivatives).

The energy density interpolation method, introduced in \cite{Masuda2012}, applies a smooth interpolation between the $\varepsilon(n_B)$ curves:
\begin{equation}
    \varepsilon(n_B) = \varepsilon_H(n_B) f_-(n_B) + \varepsilon_Q(n_B) f_+(n_B) ,
\end{equation}
where $f_\pm$ are hyperbolic tangent interpolating functions:
\begin{equation}
    f_\pm(n_B) = \frac{1}{2} \left( 1 \pm \tanh\left(\frac{n_B - \bar{n}_B}{\Gamma}\right) \right) ,
\end{equation}
with $\bar{n}_B$ and $\Gamma$ parameterizing the center and width of the phase transition. The pressure can then be calculated from the thermodynamic relation $p = n_B^2 \, \partial(\varepsilon/n_B)/\partial n_B$. This induces the following expression:
\begin{equation}
    p(n_B) = p_H(n_B) f_-(n_B) + p_Q(n_B) f_+(\rho_B) + \Delta p ,
\end{equation}
with
\begin{equation}
    \Delta p = n_B (p_H(n_B) - p_Q(n_B)) g(n_B) ,
\end{equation}
\begin{equation}
    g(n_B) = \frac{1}{2\Gamma} \cosh^{-2}\left(\frac{n_B - \bar{n}_B}{\Gamma}\right) .
\end{equation}

Both methods enable us to set the onset and length of the phase transition, which grants us additional degrees of freedom compared to a Maxwell construction. On the other hand, the applicability of these methods is also constrained to a limited range of parameters, since unphysical EoS's may also appear, where the sound speed exceeds unity or the energy density decreases with increasing chemical potential.

Throughout our investigation we will use the SFHo EoS together with the energy density interpolation as our standard choice to construct hybrid stars, although we will investigate the effect of using different hadronic EoS's and concatenation methods in Section~\ref{ssec:res_hadr_pt}.

\subsection{Compact star observables}
\label{Sec:CS_obs}

NS masses and radii can be obtained from general relativistic calculations. The line element for a spherically symmetric configuration can be expressed the following way:
\begin{equation}
    \mathrm{d}s^2 =  e^{\nu(r)}\mathrm{d}t^2 - e^{\lambda(r)}\mathrm{d}r^2 - r^2(\mathrm{d}\vartheta^2 + \sin^2\vartheta \, \mathrm{d}\varphi^2).
\end{equation}
Assuming that the matter inside NSs can be considered an approximately  spherically symmetric perfect fluid with zero temperature, and introducing the variable $m(r)$ as
\begin{equation}
e^{\lambda(r)} = \left[1-\frac{2m(r)}{r}\right]^{-1} \label{eq:tov_e} ,
\end{equation}
we can obtain the Tolman--Oppenheimer--Volkoff (TOV) equations \cite{Tolman1939,Oppenheimer1939}:
\begin{align}
\frac{\mathrm{d}m(r)}{\mathrm{d}r} &= 4\pi r^2 \varepsilon(r) , \label{eq:tov_m} \\ 
\frac{\mathrm{d}p(r)}{\mathrm{d}r} &= - [\varepsilon(r)+p(r)]\dfrac{m(r)+4\pi r^3 p(r)}{r^2 - 2 m(r) r} ,\label{eq:tov_p}
\end{align}
where $p(r)$ is the pressure related to the energy density $\varepsilon(r)$ by the EoS. Generally these equations are integrated numerically, and the boundary conditions $\varepsilon(r=0)=\varepsilon_c$, $p(R)=0$ and $m(R)=M$ determine the total mass ($M$) and radius ($R$) of the NS for a certain central energy density $\varepsilon_c$. Varying this energy density creates a sequence of NSs, and thus we obtain the $M-R$ relation for the specific EoS. Ideally one would measure the masses and radii of NSs and hence gradually constrain the $M-R$ curve and the nuclear EoS. Unfortunately NS radii are extremely difficult to measure precisely and up until now the most accurate measurements managed to achieve an accuracy of $\sim 10\%$ meaning $\sim 1-1.5$~km. The masses of NSs in binary systems, however, can be measured with remarkable precision, and in fact the most robust constraints originate from the measurements of the most massive NSs. The 2~$M_\odot$ constraint gives a powerful lower limit on the stiffness of the EoS.

Gravitational waves (GW) provide an independent way to observe NSs through their inspiral and merger with another compact object. In the final stages of the inspiral NSs are distorted through tidal interactions and this shifts the phase of the emitted GW signal. One of the measurable parameters of NS--NS mergers is the dimensionless quadrupole tidal deformability parameter $\Lambda=\lambda/M^5$, with the $\lambda$ tidal deformability being related to the $l=2$ tidal Love number:
\begin{equation}
    k_2 = \frac{3}{2} \lambda R^{-5} .
\end{equation}
Allowing small perturbations on the spherical metric, one can show that $k_2$ can be expressed the following way \cite{Hinderer2007,Damour2009}:
\begin{align}
    k_2 &= \frac{8}{5} (1-2 \beta)^2 \beta^5 [2 \beta (y_R-1)-y_R+2]\nonumber\\
    &\times \{2 \beta [4 (y_R+1) \beta^4+(6 y_R-4) \beta^3+(26-22 y_R) \beta^2\nonumber\\
    &+3 (5 y_R-8) \beta-3 y_R+6]+3 (1-2 \beta)^2\nonumber\\
    &\times[2 \beta (y_R-1)-y_R+2]\ln \left(1-2\beta\right)\}^{-1} ,
\label{eq:k2}
\end{align}
where $\beta=M/R$ is the compactness parameter of the NS and $y_R=y(R)=[rH'(r)/H(r)]_{r=R}$ with $H(r)$ being a function related to the quadrupole metric perturbation. $y_R$ can be obtained by solving the following first-order differential equation:
\begin{align}
    ry'(r)&+y(r)^2+r^2 Q(r) \nonumber\\
    &+ y(r)e^{\lambda(r)}\left[1+4\pi r^2(p(r)-\varepsilon(r))\right] = 0 ,
\label{eq:y}
\end{align}
where
\begin{align}
    Q(r)=4\pi e^{\lambda(r)}\left(5\varepsilon(r)+9p(r)+\frac{\varepsilon(r)+p(r)}{c_s^2(r)}\right) \nonumber\\
    -6\frac{e^{\lambda(r)}}{r^2}-(\nu'(r))^2 .
\label{eq:Q}
\end{align}
Here $c_s^2=\mathrm{d}p/\mathrm{d}\varepsilon$ is the sound speed squared, while $\nu'(r)$ is given by
\begin{equation}
\nu'(r) = \dfrac{2[m(r)+4\pi r^3 p(r)]}{r^2 - 2 m(r) r} \label{eq:tov_nu} .
\end{equation}

In case there is a discontinuity in the EoS (e.g. due to a first-order phase transition), owing to a $\Delta\varepsilon$ jump in the energy density at constant pressure, the term in Eq.~(\ref{eq:Q}) containing $1/c_s^2$ will diverge. One then needs to add an extra term when solving for the $y$ variable at the $r_d$ location of the discontinuity \cite{Postnikov2010,Takatsy2020}:
\begin{equation}
     y(r_d^+) - y(r_d^-) = -\frac{4\pi r_d^3 \Delta \varepsilon}{m(r_d)+4\pi r_d^3 p(r_d)} .
\label{eq:ydisc}
\end{equation}

The analysis of GW170817 inferred a value of $\Lambda<800$ for $1.4$~$M_\odot$ NSs in the low-spin limit \cite{LIGOScientific2017}. A thorough investigation of this constraint performed by Annala et al. using a generic family of EoS's found an upper radius limit of $13.6$~km for $1.4$~$M_\odot$ NSs \cite{Annala2017}. A lower radius constraint was inferred by Bauswein et al. from the absence of prompt collapse during this event \cite{Bauswein2017}, while an upper mass limit of $2.16^{+0.17}_{-0.15}$~$M_\odot$ was proposed by Rezzolla et al. using a quasi-universal relation between the maximum mass of static and uniformly rotating NSs \cite{Rezzolla2017}.

\section{Results}
\label{sec:results}

\begin{figure}[!htbp]
  \centering
  \includegraphics[width=0.48\textwidth]{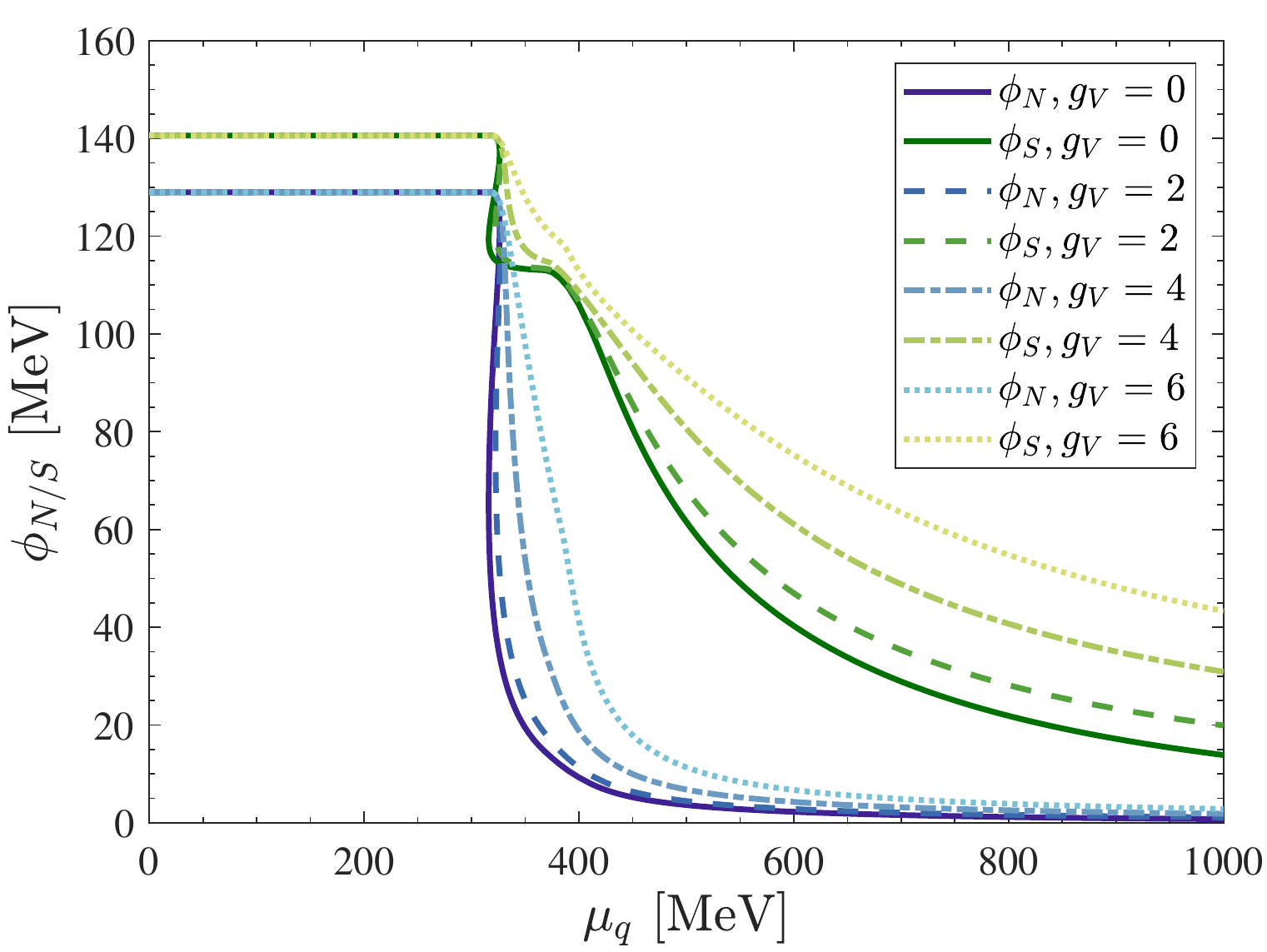}
  \caption{\label{fig:phiNS_gv}The chemical potential dependence of the $\phi_N$ (blue) and $\phi_S$ (green) scalar condensates. The parameter set with $m_\sigma=290$~MeV was used for each curve. Increasingly bright tones correspond to increasing vector couplings of $0$ (solid), $2$ (dashed), $4$ (dashed-dotted) and $6$ (dotted).}
\end{figure}

Solving simultaneously the system of six equations (Eqs.~\eqref{Eq:FE_N}-\eqref{Eq:FE_ch}) for some parameter set we get the $\mu_q$ quark chemical potential dependence of all the condensates. In Fig.~\ref{fig:phiNS_gv} the $\phi_{N/S}$ condensates are shown as a function of $\mu_q$ for different values of the $g_V$ vector coupling. For lower $g_V$ values an unstable part is present, causing a first order phase transition, which disappears for larger vector couplings.
If the phase transition is of first order at $T=0$ as a function of $\mu_B\;(=3\mu_q)$, then there is a critical end point (CEP) somewhere on the chiral phase boundary on the $T-\mu_B$ plane. Since, as it is known and will also be seen here, nonzero vector coupling is needed in order to fulfill the two solar mass criteria for the $M(R)$ curves, it seems that if $g_V\gtrsim 3.1$ (for $m_\sigma=290$~MeV) in this framework the existence of a CEP is unlikely. This is based on the observation that if in a linear sigma model the phase transition is not of first order as a function of $\mu_B$ at $T=0$, then there is no CEP on the $T-\mu_B$ plane (see e.g. \citep{Kovacs:2006ym})

\subsection{Equation of state with different vector couplings and compact star properties}

\begin{figure}[!htbp]
  \centering
  \includegraphics[width=0.48\textwidth]{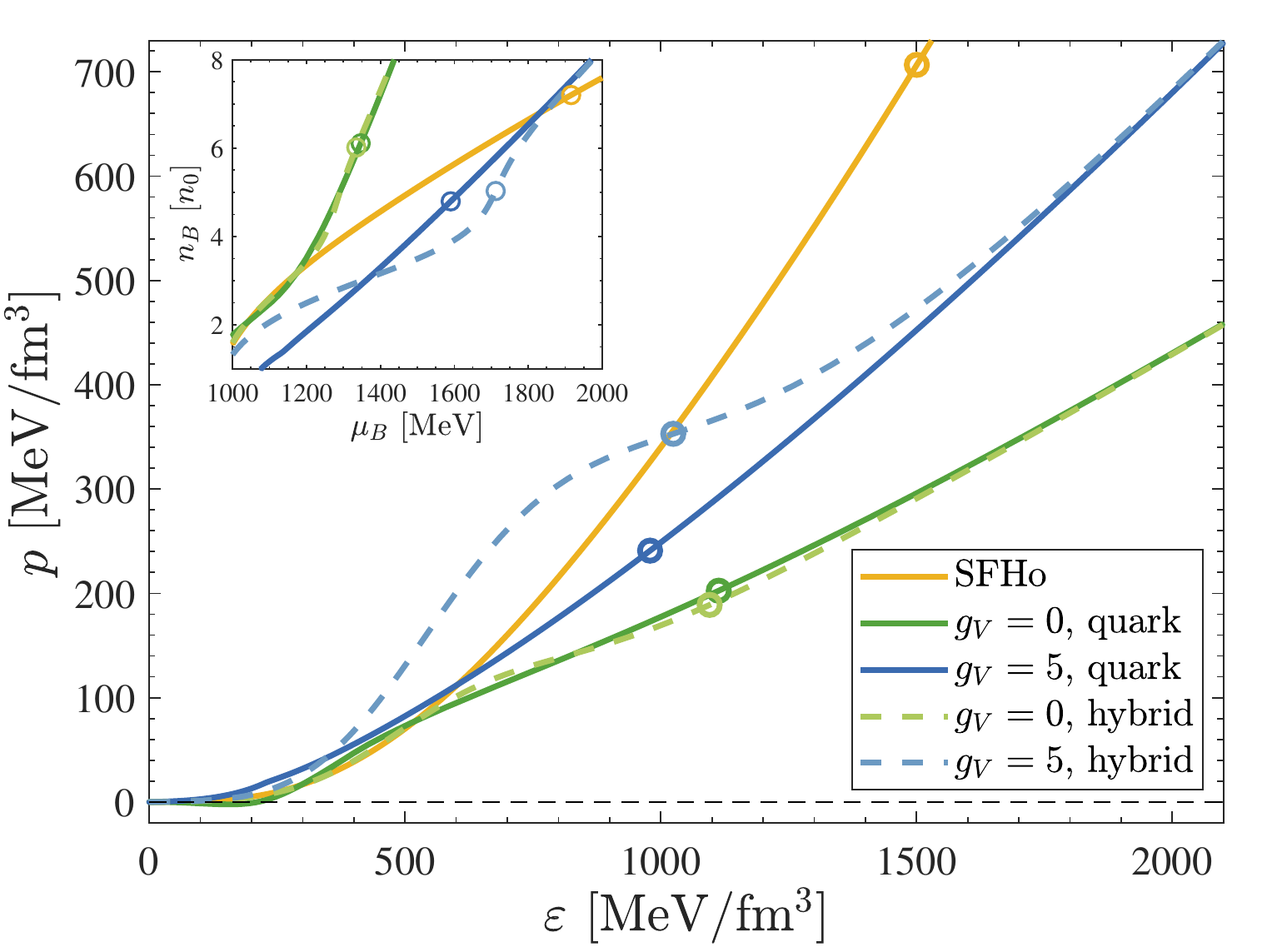}
  \includegraphics[width=0.48\textwidth]{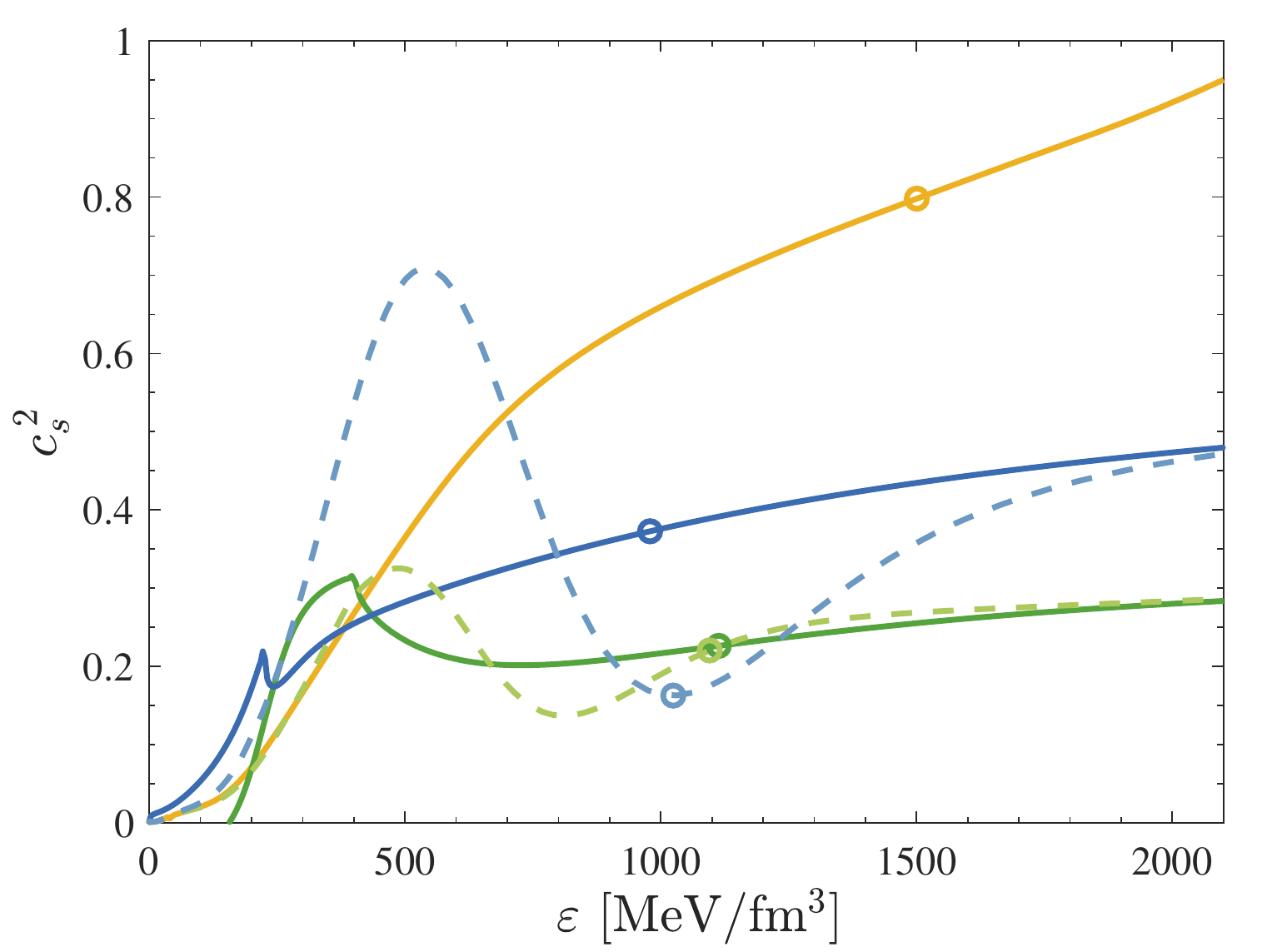}
  \caption{\textit{Top}: the EoS ($p$ as a function of $\varepsilon$) of the SFHo model (yellow solid line), as well as for quark (solid) and hybrid (dashed) stars using the eLSM with the parameter set corresponding to $m_\sigma=290$~MeV and vector couplings $g_V=0$ (green) and $g_V=5$ (blue). The inset contains the same curves for the $n_B(\mu_B)$ dependence, while the circles correspond to the central conditions inside the maximum mass neutron stars. For the hybrid EoS's the energy density interpolation was used with $\bar{n}_B=3.5n_0$ and $\Gamma=1.5n_0$. \textit{Bottom}: the speed of sound squared for the same EoS's.}
\label{fig:EoS_gv}
\end{figure}

\begin{figure}[!htbp]
  \centering
  \includegraphics[width=0.48\textwidth]{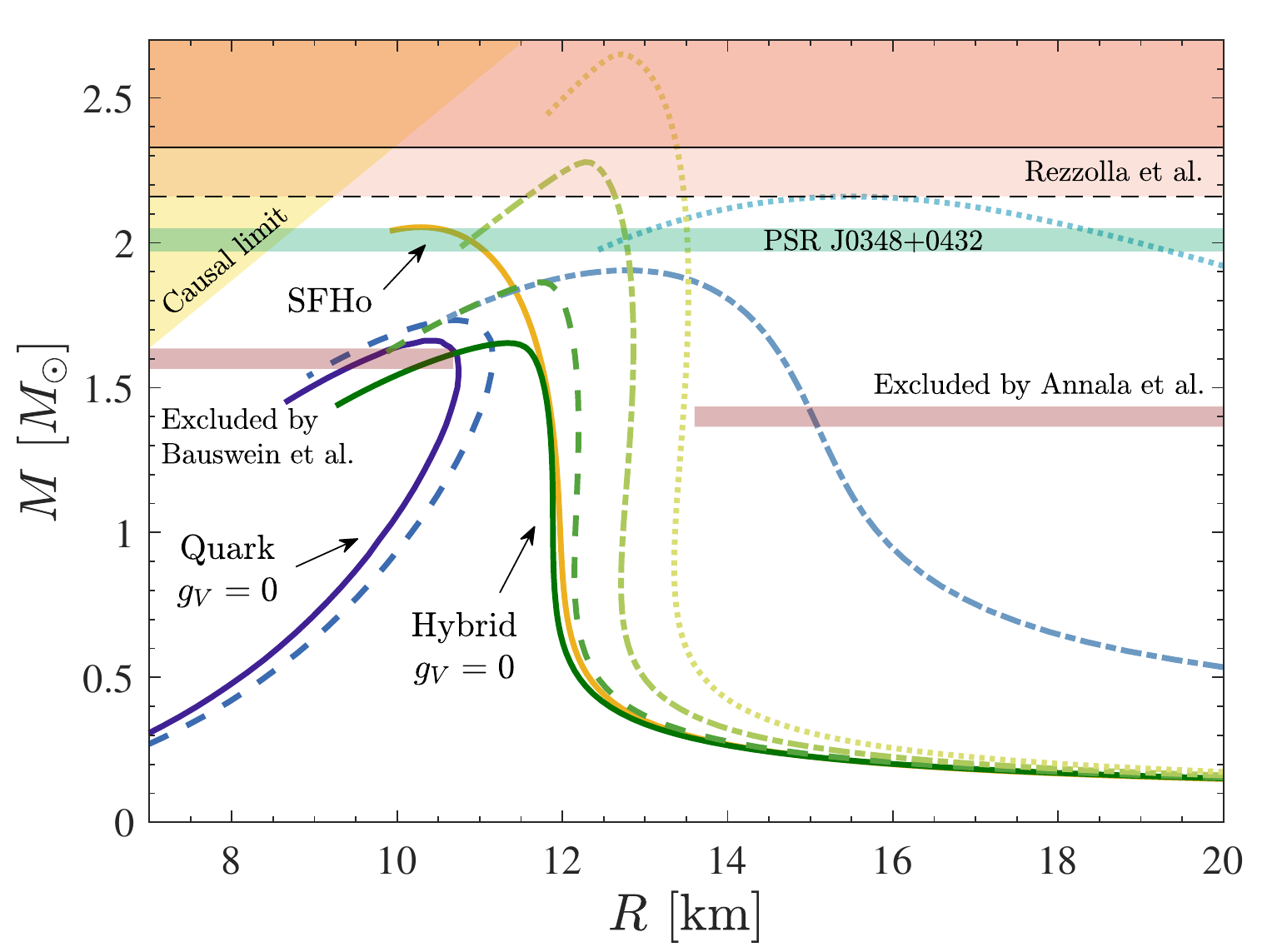}
  \caption{$M-R$ relations for the SFHo model (yellow) and for different quark (blue) and hybrid (green) EoS's. Increasingly bright tones correspond to increasing vector couplings of $0$ (solid), $2$ (dashed), $4$ (dashed-dotted) and $6$ (dotted). The different shaded areas correspond to the lower and upper radius constraints of Bauswein \textit{et al.} \cite{Bauswein2017} and Annala \textit{et al.} \cite{Annala2017}, respectively, different credibility limits of the upper mass constraint of Rezzolla \textit{et al.} \cite{Rezzolla2017}, and the mass of PSR J0348+0432 \cite{Antoniadis2013}. The yellow shaded area corresponds to the region excluded by causality.}
\label{fig:MR_gv}
\end{figure}

Along the solution (of Eqs.~\eqref{Eq:FE_N}-\eqref{Eq:FE_ch}) the pressure $p$ (Eq.~\eqref{Eq:pressure}) and its derivatives can be calculated, from which one can construct the equation of state (EoS). The EoS, which is the pressure as the function of the energy density $\varepsilon$, can be seen in Fig.~\ref{fig:EoS_gv} for the SFHo hadronic model, for the pure quark and for the hybrid stars for two different values of $g_V$. 
For the case $g_V=0$ the hybrid EoS smoothly connects the hadronic and quark phases. However, for the case $g_V=5$, even though the quark EoS is softer than the hadronic one, an intermediate region appears where the hybrid EoS becomes stiffer than both the quark and the hadronic ones. This results in an increase of the maximum compact star mass (see Fig.~\ref{fig:MR_gv}). We note that this behaviour is not the consequence of the specific choice for the concatenation method, since it appears for the $p(\mu)$ interpolation as well (see the comparison of the concatenation methods in \ref{ssec:res_hadr_pt}). The same results were reported already in \cite{Masuda2012}, while other studies investigating a hadron--quark continuity also found similar results \cite{Kojo2021}.

As described in Sec.~\ref{Sec:CS_obs} the EoS's are needed to calculate the $M-R$ curves and $\Lambda$ tidal deformability parameters for a compact star. In Fig.~\ref{fig:MR_gv} the $M-R$ curves can be seen for the EoS's of the SFHo model, of the pure quark model and of hybrid stars with various $g_V$ vector couplings. 
We see that while the low-density (low-mass) behaviour of the relations for hybrid stars is determined by the hadronic EoS, the maximum mass region is characterized by the quark EoS. For $g_V=0$, where the hybrid EoS smoothly interpolates between the two phases (see Fig.~\ref{fig:EoS_gv}), the quark and hybrid models describe maximum mass compact stars with approximately the same masses. For larger vector couplings, on the other hand, the maximum mass is greatly increased due to the intermediate stiffening of the hybrid EoS. The radii of quark stars with larger vector couplings are also greatly increased owing to the absence of the first order phase transition and an incorrect low-density behaviour -- due to the lack of proper degrees of freedom, i.e. the baryons, at low densities.

\begin{figure}[!htbp]
  \centering
  \includegraphics[width=0.48\textwidth]{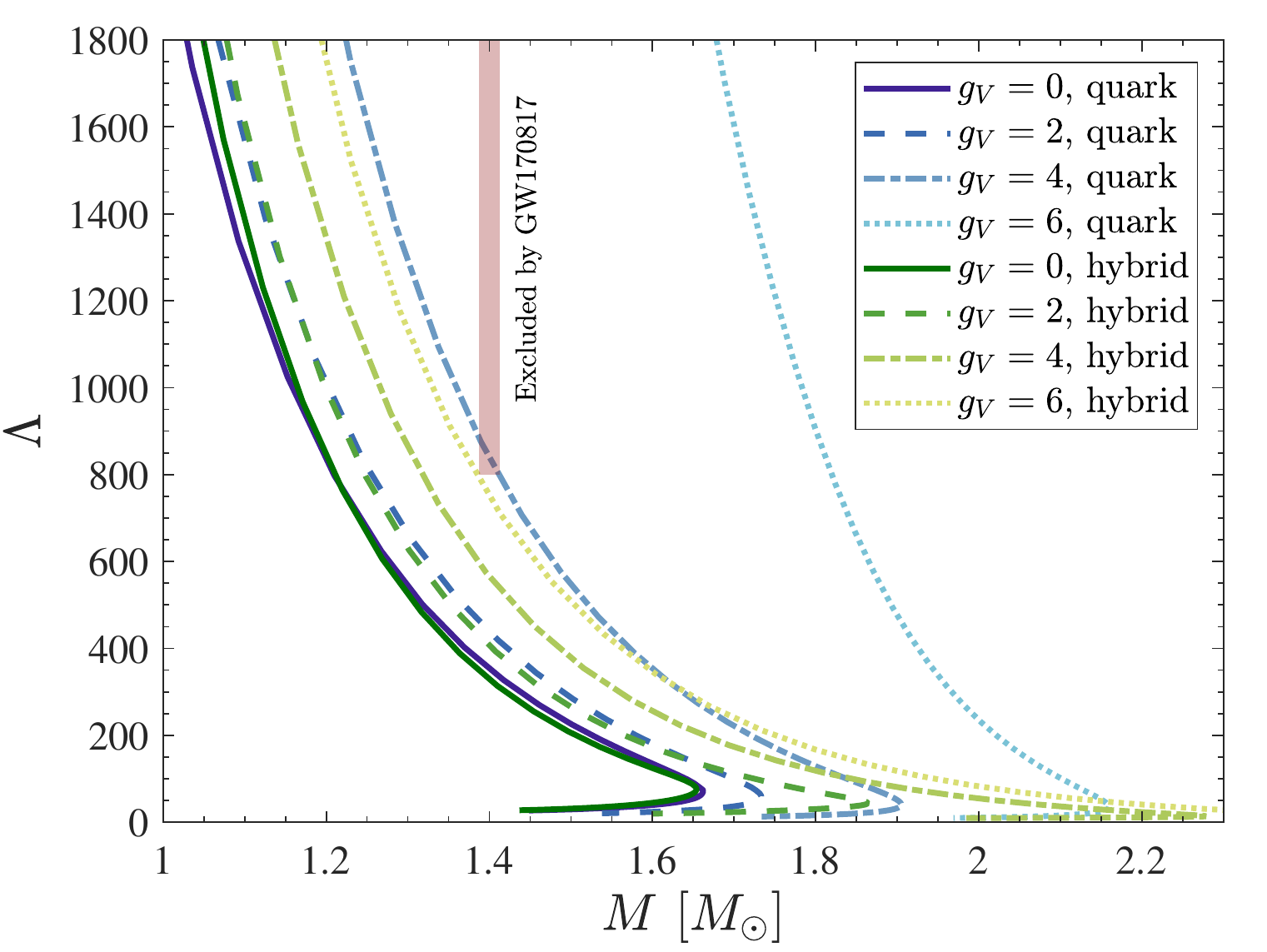}
  \caption{The same curves as in Fig.~\ref{fig:MR_gv}, but for the tidal deformabilities $\Lambda$ as function of the compact star masses $M$. The red bar corresponds to the region excluded by the constraint $\Lambda(1.4M_\odot)<800$ deduced from the measurement of GW170817 \cite{LIGOScientific2017}.}
\label{fig:Lambda_gv}
\end{figure}

In Fig.~\ref{fig:Lambda_gv} the tidal deformability parameter versus the compact star mass can be seen for the same EoS's as in the case of Fig.~\ref{fig:MR_gv}. 
\begin{figure}[!htbp]
  \centering
  \includegraphics[width=0.48\textwidth]{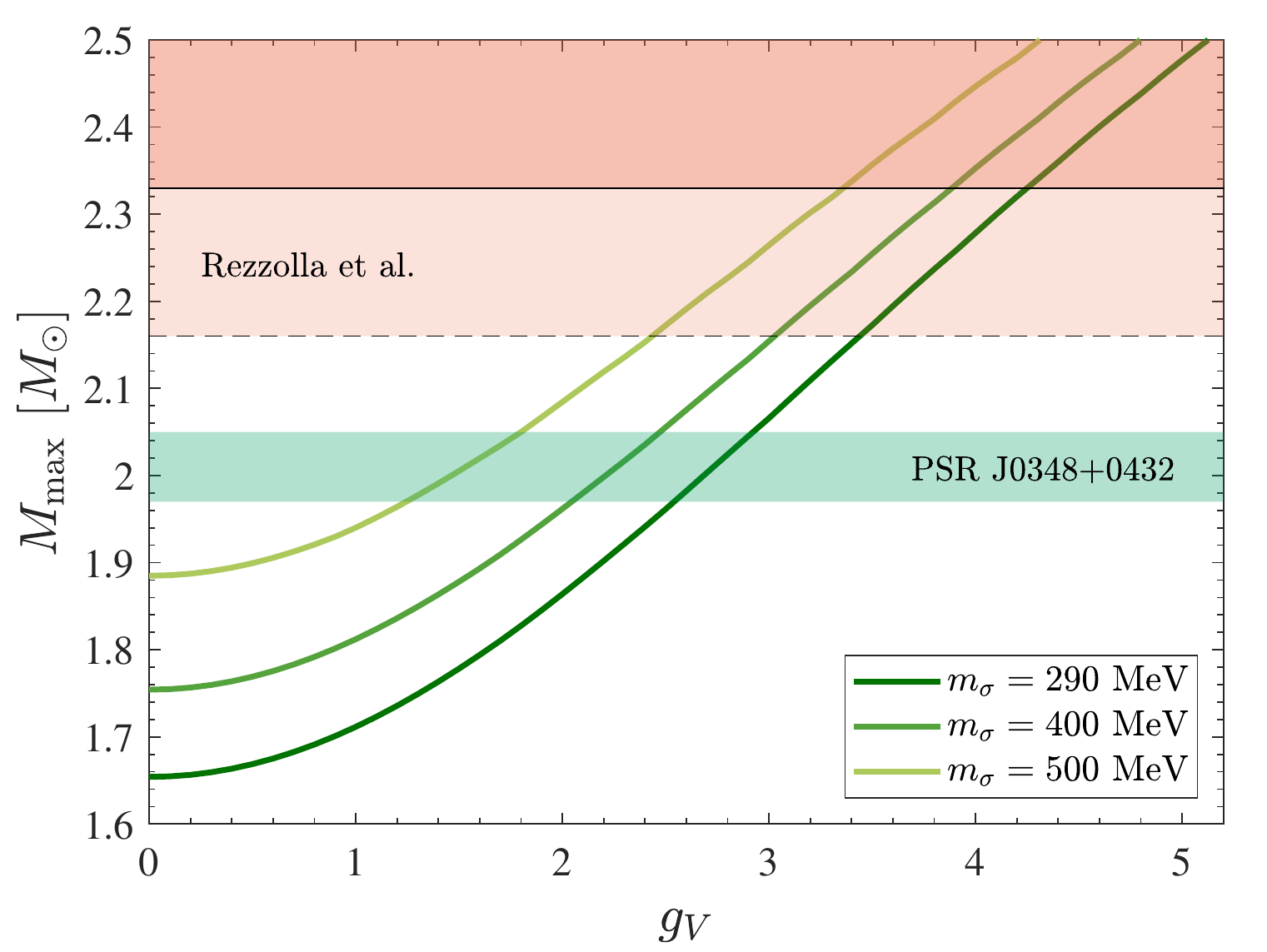}
  \caption{The maximum mass of stable neutron star sequences as a function of the vector coupling $g_V$, for different sigma masses. The parameters of the concatenation are $\bar{n}=3.5n_0$ and $\Gamma=1.5n_0$. For higher sigma masses this results in EoS's that violate either thermodynamic stability or causality.}
\label{fig:Mmax_gv}
\end{figure}
Note that the excluded region in the figure corresponds to the upper radius constraint of Annala \textit{et al.} \cite{Annala2017} on the $M-R$ plot, and it excludes the same models. Currently the measurements of the tidal deformabilities can not provide stricter constraint on the EoS's than mass and radius measurements, however, both are expected to be improved with upcoming measurements. From the two figures we can conclude that considering hybrid stars the vector coupling should be in the following range to meet all the requirements,
\begin{equation}
    2.6<g_V<4.3.
\end{equation}
This range is valid in case of $m_{\sigma} = 290$~MeV and can be directly seen in Fig. \ref{fig:Mmax_gv}, where the maximum mass of stable neutron stars is shown as a function of $g_V$ for different $m_{\sigma}$. The lower value of the range is given by the intersection point of the curve with bottom of the green band (titled PSR J0348+0432), while the upper one is from the intersection point of the curve with the top of the light rose-colored band (titled Rezzola et al.) 

\subsection{Dependence on the sigma meson mass}

\begin{figure}[!htbp]
  \centering
  \includegraphics[width=0.48\textwidth]{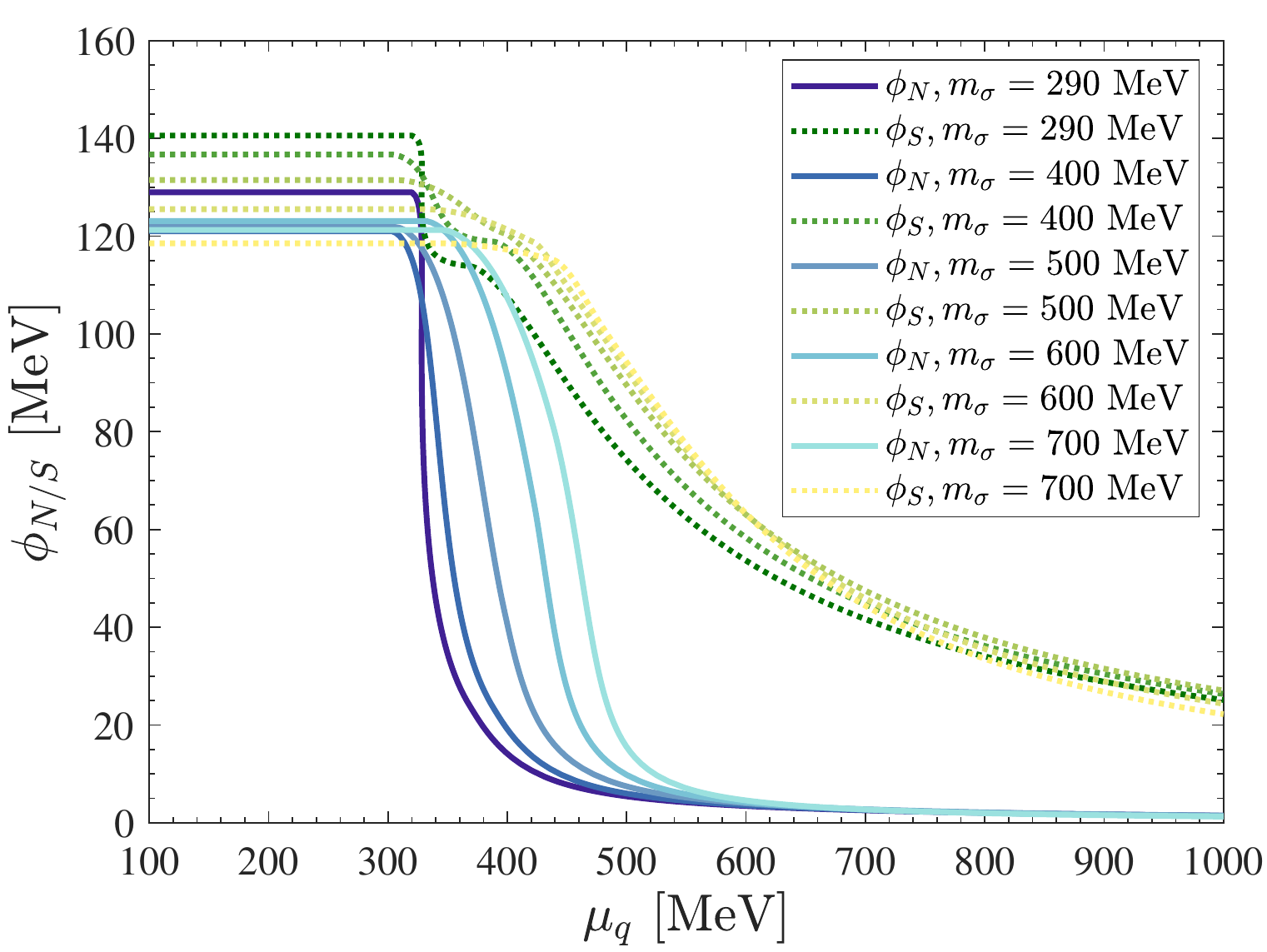}
  \caption{The chemical potential dependence of the $\phi_N$ (blue solid) and $\phi_S$ (green dotted) scalar condensates. Increasingly bright tones correspond to parameter sets with increasing sigma meson masses with $g_V=3$ for each curve. For larger $m_\sigma$ masses the first order phase transition (indicated by the slight back-bending of the $\phi(\mu_q)$ curves) disappears, and for $m_\sigma=700$ MeV the vacuum expectation value of $\phi_N$ becomes larger than the value of $\phi_S$.}
\label{fig:PhiNS_msigma}
\end{figure}

Beside $g_V$ there is another very important parameter, the mass of the $f_0$ or $\sigma$ meson, that substantially changes the behavior of the solution to the field equations and consequently the behavior of the EoS itself. Its very important role comes from the fact that the $\sigma_N$ non-strange and the $\sigma_S$ strange scalar fields acquire nonzero condensates -- which are the $\phi_N$ and $\phi_S$ -- in the meson sector. However, it is worth noting that other condensates, like pion or kaon condensates, are also considered in the literature, see e.g. \cite{Herpay:2008uw, Thorsson:1993bu}, but that is out of the scope of the current investigation. In Fig.~\ref{fig:PhiNS_msigma} one can see the $\phi_{N/S}$ condensates as a function of the $\mu_q$ quark chemical potential for different values of $m_{\sigma}$. 
As it can be seen, the phase transition is first order only for very low values of $m_{\sigma}$ , which is indicated by the slight back-bending of the $m_{\sigma}=290$~Mev curve for $\phi_{N/S}$. If $m_{\sigma}\gtrsim 300$~MeV the transition becomes crossover and the pseudocritical chemical potential -- defined by the inflection point of the $\phi_N(\mu_q)$ curve -- shifts toward larger values. 

\begin{figure}[!htbp]
  \centering
  \includegraphics[width=0.48\textwidth]{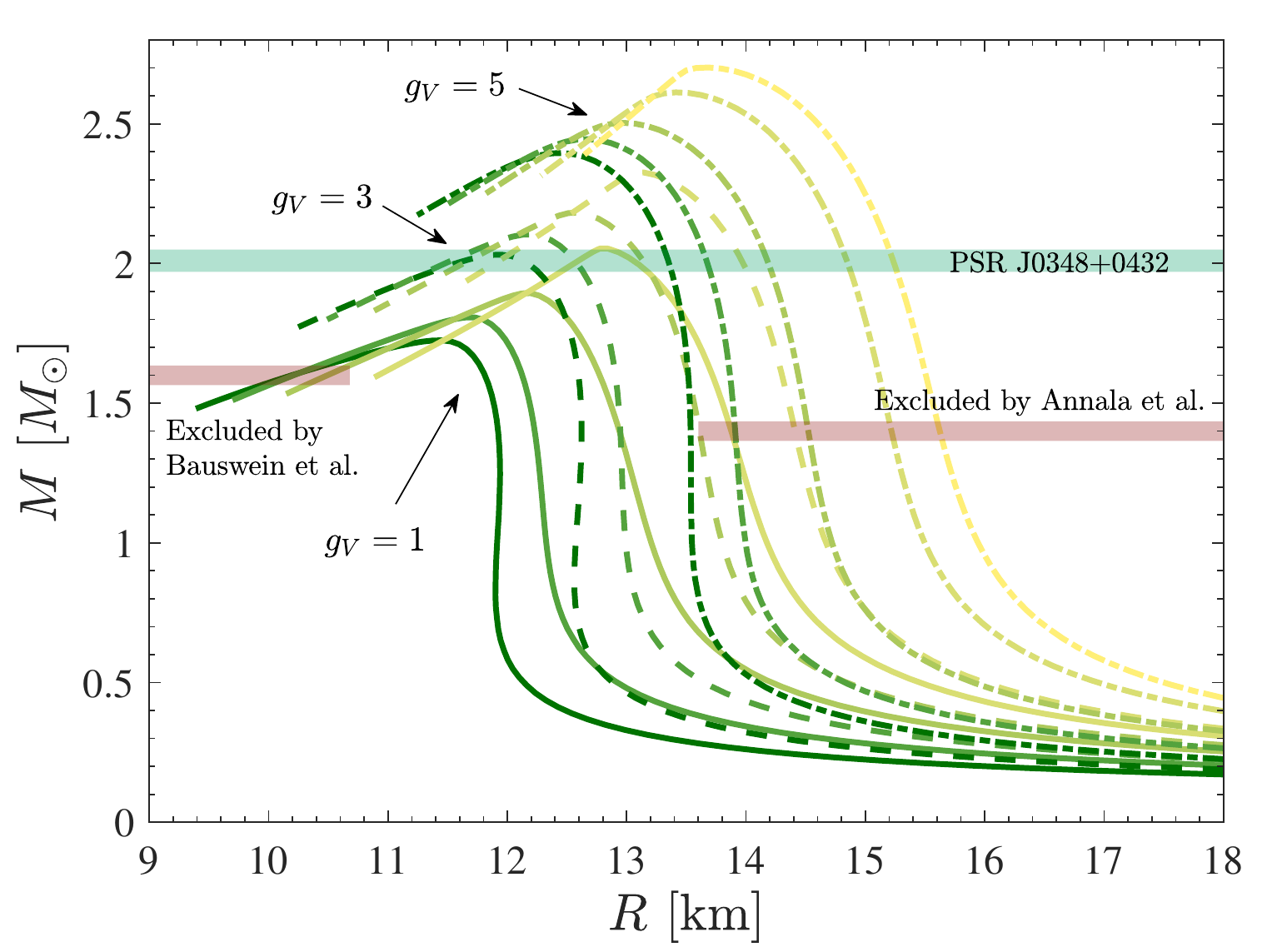}
  \caption{$M-R$ relations of hybrid stars for different vector couplings and sigma meson masses, using the SFHo hadronic EoS and the energy density interpolation with $\bar{n}_B=3.5n_0$ and $\Gamma=2n_0$. Increasingly bright tones correspond to the five parameter sets with increasing $m_\sigma$ (the same as in Fig.~\ref{fig:PhiNS_msigma}), while the different types of lines correspond to different vector couplings of $1$ (solid), $3$ (dashed) and $5$ (dashed-dotted). The curves for $m_\sigma=700$~MeV with $g_V=1$ and $3$ are omitted, since the hybrid EoS's produced with these parameters are not stable.}
\label{fig:MR_msigma}
\end{figure}

The effect of changing $m_{\sigma}$ for three different $g_V$ values can be seen in Fig.~\ref{fig:MR_msigma}.  
It can be observed that generally the larger $m_\sigma$ and $g_V$ are, the larger the compact star masses and radii are. However, the value of $m_\sigma$ moderately modifies the slope of the $M-R$ curve in the mid-mass region. For a given $g_V$ value the change in the value of the maximum mass is about $15-17\%$ for the total range of the $\sigma$ mass studied here. On the other hand, if for a fixed $m_{\sigma}$ we change $g_V$ from 1 to 5 we get an approximately $40\%$ change in the maximum mass.

\subsection{Dependence on hadronic EoS and concatenation method and the role of the bag constant}
\label{ssec:res_hadr_pt}

\begin{figure}[!htbp]
  \centering
  \includegraphics[width=0.48\textwidth]{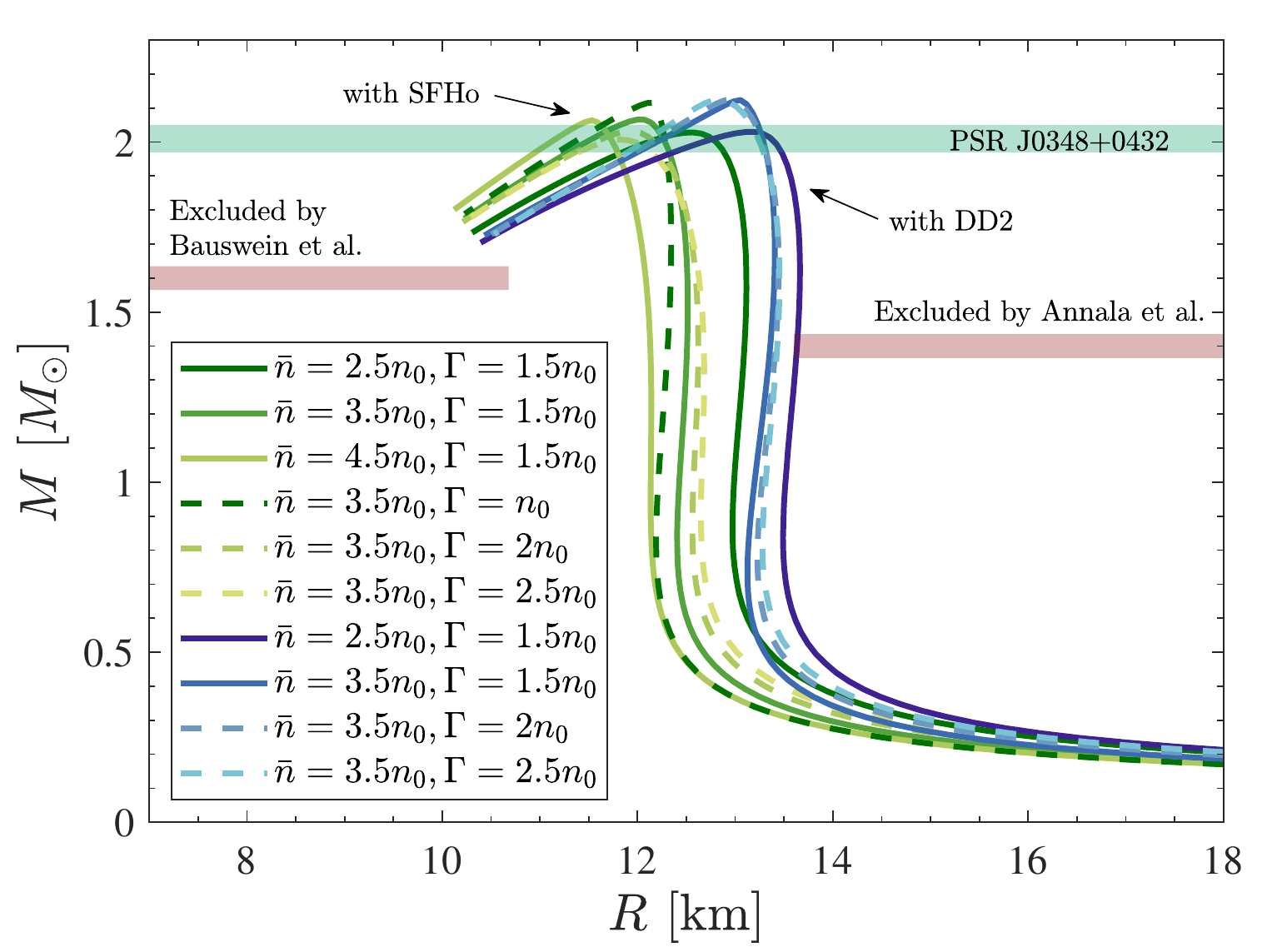}
  \caption{$M-R$ relations of hybrid stars with the SFHo (green) and the DD2 (blue) hadronic EoS's and the energy density interpolation method with different values for $\bar{n}_B$ (solid) and for $\Gamma$ (dashed). Brighter tones correspond to larger values in both cases. $g_V=3$ for each curve. For the DD2 case only the $M-R$ curves of stable EoS's are included. }
\label{fig:MR_hadr}
\end{figure}

We also investigated the effects of changing the hadronic EoS and the method of concatenation. In Fig.~\ref{fig:MR_hadr} the $M-R$ curves are shown for different $\bar{n}_B$ and $\Gamma$ values with the SFHo and DD2 hadronic models. 
Even though the radii of hybrid stars are dependent on the choice of the phase transition parameters and the hadronic EoS, the maximum mass allowed by the different models is encompassed within a small range.

\begin{figure}[!htbp]
  \centering
  \includegraphics[width=0.48\textwidth]{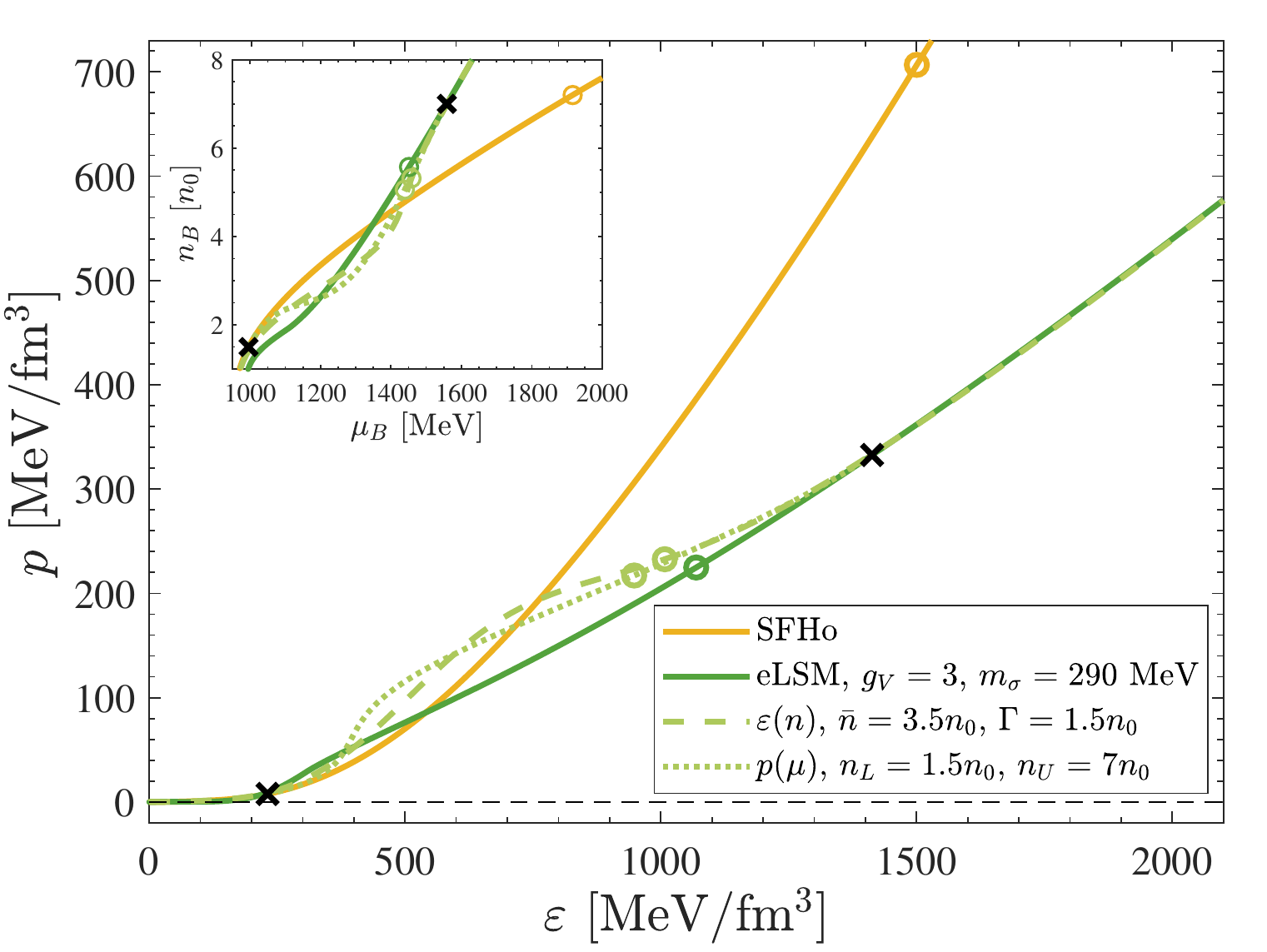}
  \includegraphics[width=0.48\textwidth]{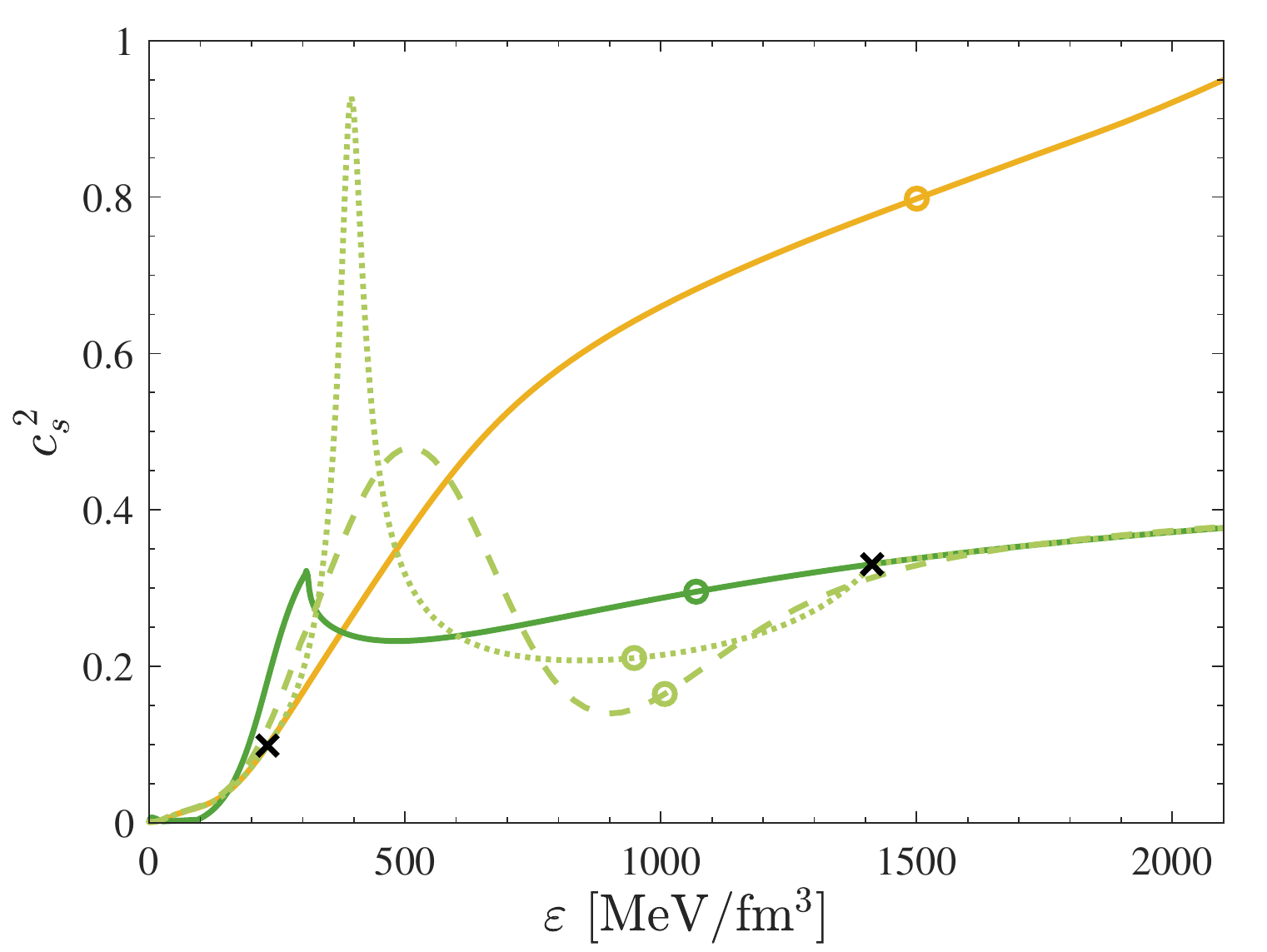}
  \caption{\textit{Top}: hybrid EoS's produced by the two concatenation methods using the SFHo model (yellow solid line) and the eLSM with the parameter set corresponding to $m_\sigma=290$~MeV and $g_V=3$ (green solid line). The black crosses correspond to $n_L\equiv n_{B,H}(\mu_{BL})$ and $n_U\equiv n_{B,Q}(\mu_{BU})$, while the circles correspond to the central conditions inside the maximum mass neutron stars. The parameters of the two types of interpolation methods were chosen so that both arrive to the hadronic and quark EoS's at approximately the same densities. \textit{Bottom}: the speed of sound squared for the same EoS's.}
\label{fig:EoS_pt}
\end{figure}

As it is discussed in Sec.~\ref{ssec:hq_trans} two different kinds of interpolation were used in our investigations, one that interpolates between the energy densities as a function of the baryon density and another one that uses the pressure as a function of the baryochemical potential. The comparison of the two approaches can be seen in Fig.~\ref{fig:EoS_pt} for given values of $g_V$ and $m_{\sigma}$.
The $\varepsilon(n)$ (dashed) and the $p(\mu)$ (dotted) interpolations both show similar features with stiffenings in the intermediate-density region, although this starts at lower densities for the $p(\mu)$ case. Even if the two kind of interpolation methods use very different functions -- a polynomial and a tangent hyperbolic -- we see similar behavior in the intermediate region. It is also worth noting that, even though the EoS's look similar, more pronounced difference in the speed of sound are apparent for the two different interpolation methods.
\begin{figure}[!htbp]
  \centering
  \includegraphics[width=0.48\textwidth]{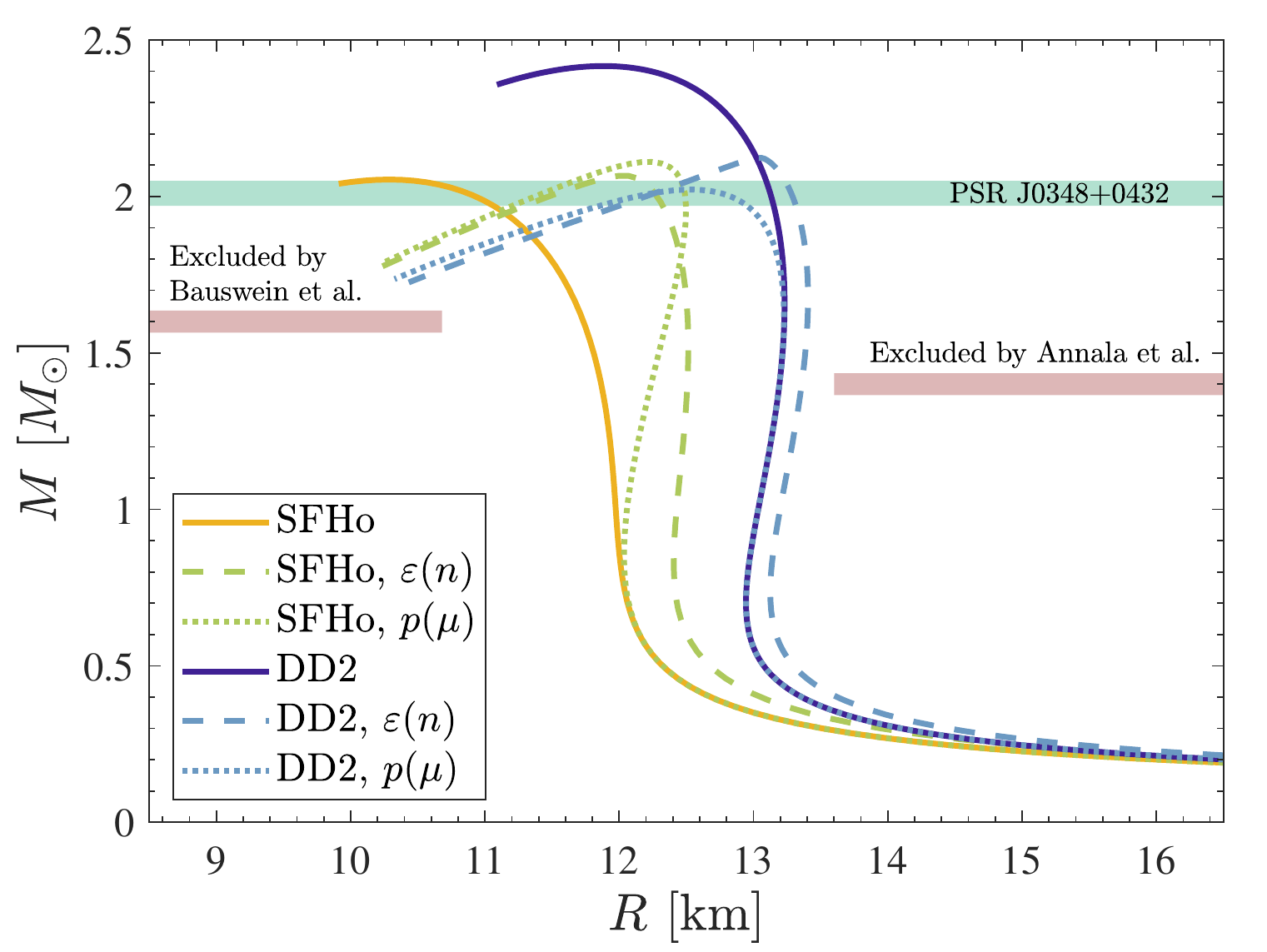}
  \caption{$M-R$ relations of hybrid stars for different concatenations as well as for the SFHo (yellow) and DD2 (blue) hadronic EoS's. The parameters for the $\varepsilon(n)$ (dashed) and $p(\mu)$ (dotted) concatenations are the same as in Fig.~\ref{fig:EoS_pt}.}
\label{fig:MR_pt}
\end{figure}

In Fig. \ref{fig:MR_pt} the effect of the different interpolation methods are shown for the two types of hadronic EoS's, the SFHo and the DD2. In both cases there is a slight change in the values of the maximal mass that happens oppositely for the two hadronic EoS's. In the middle mass range there is also a slight change in the radii, which act in the same way for the two hadronic curves by increasing the radius, which shows a smaller effect  for the case of DD2 . With the current observations neither of the scenarios depicted here can be excluded.

Finally, we have also investigated the role of the bag constant $B$ in the current framework. We have taken different values for $B^{1/4}$ from $0$ to $110$~MeV (similarly to \cite{Zacchi:2019ayh}). The bag constant represents a vacuum contribution and it is simply an additional constant for the pressure and the energy density.
\begin{figure}[!htbp]
  \centering
  \includegraphics[width=0.48\textwidth]{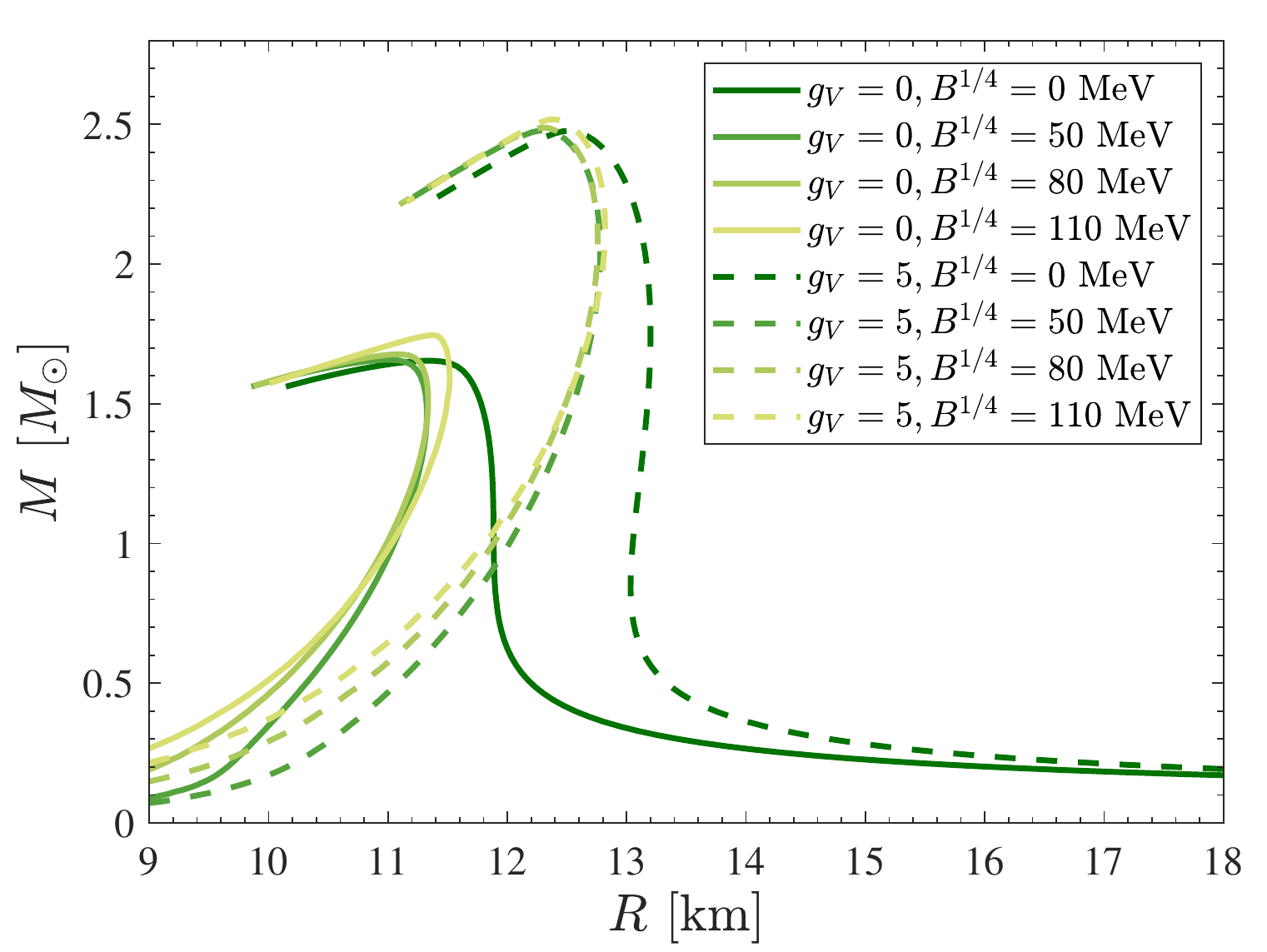}
  \includegraphics[width=0.48\textwidth]{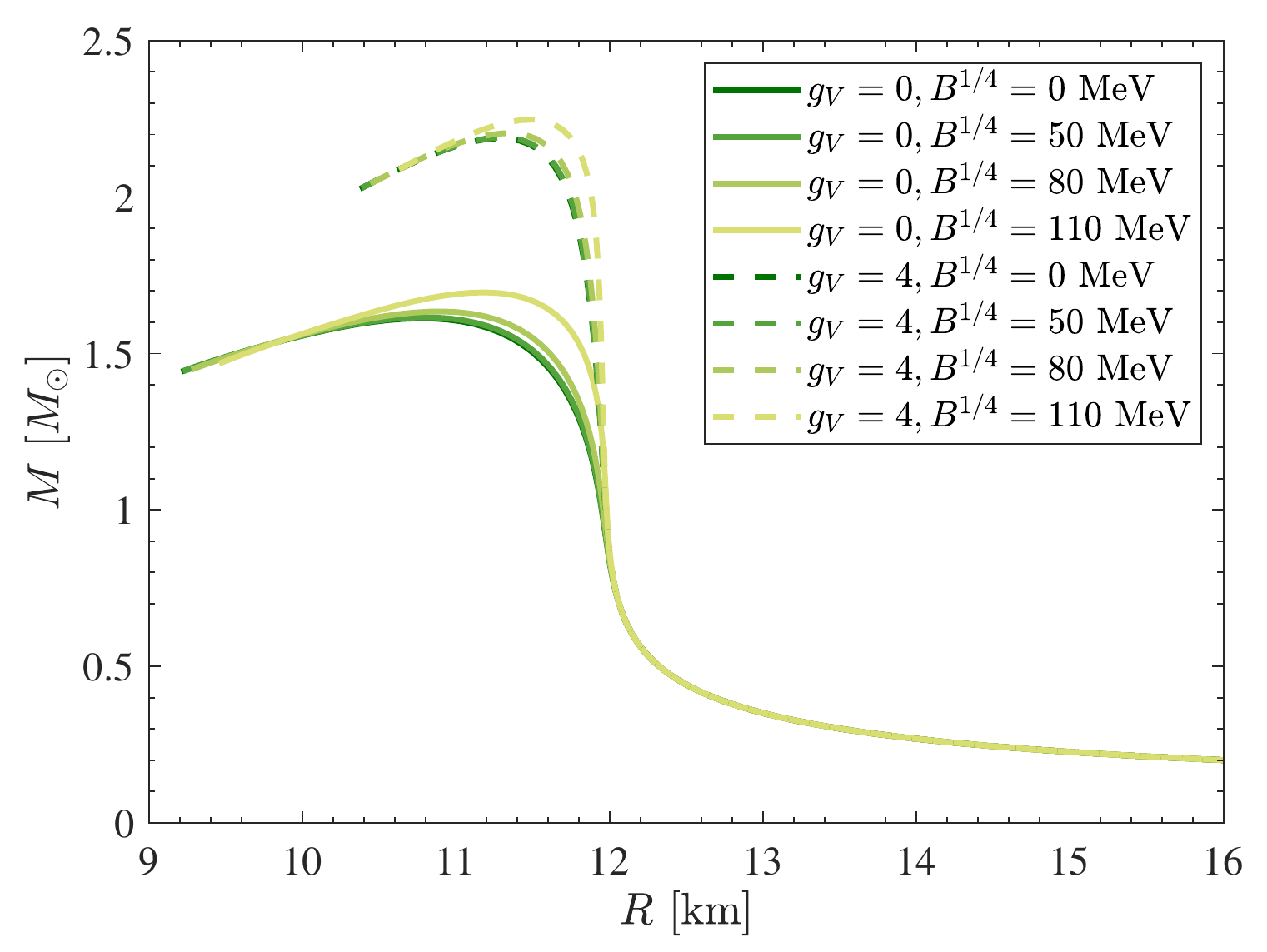}
  \caption{$M-R$ relations in case of a nonzero bag constant for the $\varepsilon(n)$ (top) and the $p(\mu)$ (bottom) interpolation methods. In the $p(\mu)$ case the interpolation limits were chosen so that larger values for $B$ could be accommodated as well without producing unstable EoS's.}
\label{fig:MR_B}
\end{figure}
Many previous studies have also investigated the effect of changing the bag constant only using a Maxwell construction for modeling the phase transition (e.g. \cite{Zacchi:2019ayh,Cierniak2020}). Some of these studies found that the M--R curves cross each other in the vicinity of a single point, the so-called "special point" \cite{Yudin2014,Cierniak2020,Cierniak2021}. In our case, due to the differing nature of our phase transition construction, we do not expect such a special point to appear.

In Fig. \ref{fig:MR_B} the effect of a nonzero $B$ term can be seen for the two kind of interpolation methods. On the upper figure the $M-R$ relations for the $\varepsilon(n)$ concatenation are shown for two $g_V$ and four $B^{1/4}$ values. As $B^{1/4}$ increases, the low mass neutron stars will develop small radii -- as if they were pure quark stars -- due to the incorrect low-density behaviour. This behavior can not be observed in case of the $p(\mu_B)$ concatenation in the lower figure. Beside this change the effect of the bag constant is not so dramatic in our case.

\section{Conclusions}
\label{sec:conclusion}

We investigated hybrid star properties using the concept of hadron-quark crossover, in which we took a hadronic EoS together with a quark one and connected them with some smooth interpolation method in an intermediate region where both models are inaccurate. For the hadronic EoS we have used two different relativistic mean field models, the SFHo and the DD2, while for the quark part an (axial)vector meson extended linear sigma model was used with additional constituent quarks. The latter model reproduces the meson spectrum in vacuum well and also agrees with various lattice results at finite temperature and zero density.

We argued that the changes in the values of the parameters of the Lagrangian can have a significant effect on the properties of the EoS and consequently on the properties of hybrid stars themselves. For this very reason, we investigated the asymptotic behavior of our system of equations as a function of the $\mu_q$ quark chemical potential and found a condition among a set of parameters of our Lagrangian that should be fulfilled in order to acquire vanishing chiral condensates for very large $\mu_q$, as it is expected physically.

The interpolation method was also altered and its effect on the $M-R$ curves was shown to be moderate in the mid-mass range, while the maximum hybrid star mass remained approximately constant. Moreover the parameters of the interpolation -- like position and width --, the $g_V$ vector coupling and the $m_{\sigma}$ sigma meson mass were also changed in some range and their effect were analyzed in detail. We found that for a given value of the sigma meson mass there is a relatively small acceptable range in $g_V$, imposing constraints from astrophysical observations. The consequences of the constraints on the tidal deformabilities for different EoS's and $g_V$ vector couplings was also discussed. Finally we found in connection with the bag constant that its introduction does not affect the $M-R$ curves significantly if its value is not too high for a given $g_V$ value.

In conclusion, all the current astrophysical constraints from observation are compatible  with the investigated phenomenological model if the relevant parameters -- like the vector coupling or the sigma meson masses -- are within a certain range and the parameters also satisfy a condition that comes from the investigation of the asymptotic behavior of the field equations. Turning the argument around, one sees that data from neutron stars considerably constrain the parameters of the chiral quark-meson model for bulk quark matter.

It is worth to note that the $g_V$ vector coupling can also be determined from the parameterization procedure if one uses one-loop order curvature masses for the vector and axial vector meson masses \cite{Kovacs:2021kas}. In \cite{Kovacs:2021kas} its value was found to be around $5$ for a sigma mass around $300$~MeV, which is a little higher than the upper bound of the acceptable range found here. To resolve this tension further investigation is needed.

\begin{acknowledgements}
J. T. and P. K. acknowledge support by the National Research, Development and Innovation (NRDI) fund of Hungary, financed under the FK\_19 funding scheme, Project No. FK 131982. J. T., P. K. and Gy. W. acknowledge the support of the NRDI fund K138277. P. K. also acknowledges support by the János Bolyai Research Scholarship of the Hungarian Academy of Sciences. J. T. was supported by the ÚNKP-21-3 New National Excellence Program of the Ministry for Innovation and Technology from the source of NRDI fund.
J. S.-B. acknowledges support by the Deutsche Forschungsgemeinschaft (DFG, German Research Foundation) – project number 315477589 – TRR 211.
\end{acknowledgements}

\appendix

\section{Grand potential in the hybrid approximation}
\label{app:omega}

The grand potential in the current (hybrid) approximation reads
\begin{widetext}
\begin{align}
    \Omega_{\textrm{tot}} &= \frac{1}{2}m_0^2 \left(\phi_N^2 + \phi_S^2 \right) +  \frac{\lambda_1}{4}\left(\phi_N^2 + \phi_S^2 \right)^2 + \frac{\lambda_2}{8}\left(\phi_N^4 + 2\phi_S^4 \right) -\frac{c_1}{2\sqrt{2}}\phi_N^2\phi_S - h_N\phi_N - h_S\phi_S  \nom \\ &- \frac{1}{2}m_{\rho}^2\left(v_{\omega}^2 + v_{\rho}^2 \right) - \frac{1}{2}m_{\Phi}^2v_{\Phi}^2
    -\frac{3}{8\pi^2}\left[ m_u^4 \log\left(\frac{m_u}{M_0}\right) + m_d^4 \log\left(\frac{m_d}{M_0}\right) + m_s^4 \log\left(\frac{m_s}{M_0}\right) \right] \nom \\
    &- \frac{3}{8\pi^2}\sum_{f\in (u,d,s)}m_f^4\left[\log\left( \gamma_f + \sqrt{\gamma_f^2 - 1} \right) + \frac{1}{3}\gamma_f \sqrt{\gamma_f^2 - 1 } \left( 2 \gamma_f^2 - 5 \right)\right] \nom \\
    &- \frac{m_e^4}{8\pi^2} \left[\log\left( \gamma_e + \sqrt{\gamma_e^2 - 1} \right) + \frac{1}{3}\gamma_e \sqrt{\gamma_e^2 - 1 } \left( 2 \gamma_e^2 - 5 \right)\right]
    \label{Eq:GP}
\end{align}
where the vector masses are given by
\begin{align}
    m_{\rho}^2 &= m_{\omega}^2 = m_1^2 + \frac{1}{2}h_1\left(\phi_N^2 + \phi_S^2 \right) + \frac{1}{2}\left(h_2 + h_3\right)\phi_N^2, \label{Eq:mrho}\\
    m_{\Phi}^2 &= m_1^2 + \frac{1}{2}h_1\left(\phi_N^2 + \phi_S^2 \right) + \left(h_2 + h_3\right)\phi_S^2 + 2\delta_S,\label{Eq:mphi}
\end{align}
and
\begin{equation}
\label{Eq:gamma_f}
    \gamma_f = \frac{|\Tilde{\mu}_f|}{m_f},\, f\in(u,d,s),\quad \gamma_e = \frac{|\mu_e|}{m_e} 
\end{equation}
have also been introduced.
\end{widetext}

\section{Parameter sets}
\label{app:param}

In Table~\ref{Tab:param_sets} we present all the parameter sets that were used for the different sigma masses.
\begin{widetext}
\begin{table*}[hbt!]
\caption{Parameter values for different sigma meson masses\label{Tab:param_sets}}
\centering
\begin{tabular}[c]{|c|c|c|c|c|}\hline
 Parameter & $m_\sigma=400$ MeV & $m_\sigma=500$ MeV & $m_\sigma=600$ MeV & $m_\sigma=700$ MeV \\\hline\hline
 $\phi_{N}$ [GeV] & 0.1210 & 0.1218 & 0.1231 & 0.1212 \\\hline
 $\phi_{S}$ [GeV] & 0.1367 & 0.1315 & 0.1255 & 0.1185 \\\hline
 $m_{0}^2$ [GeV$^2$] & -3.5902\e{-2} & -8.6251\e{-2} & -0.1780 & -0.2777 \\\hline
 $m_{1}^2$ [GeV$^2$] & 0.5600 & 0.5600 & 0.5600 & 0.5600 \\\hline
 $\lambda_{1}$ & 0.1757 & 0.9634 & 1.3876 & 2.3572 \\\hline
 $\lambda_{2}$ & 23.0159 & 25.9249 & 29.7608 & 34.9895 \\\hline
 $c_{1}$ [GeV] & 1.5884 & 1.6273 & 1.6407 & 1.7110 \\\hline
 $\delta_{S}$ [GeV$^2$] & 0.2371 & 0.2323 & 0.2305 & 0.2351 \\\hline
 $g_{1}$ & 5.3250 & 5.4668 & 5.6116 & 5.8237 \\\hline
 $g_{2}$ & -2.5367 & -2.1499 & -1.6939 & -1.4069 \\\hline
 $h_{1}$ & 5.3887 & 3.9904 & 0.6246 & -1.1022 \\\hline
 $h_{2}$ & -10.3257 & -7.4554 & -1.9978 & 1.4249 \\\hline
 $h_{3}$ & 1.8646 & 1.4698 & 1.0551 & 0.2283 \\\hline
 $g_{F}$ & 5.0050 & 5.0396 & 5.3835 & 5.7157 \\\hline
 $M_{0}$ [GeV] & 0.4522 & 0.4522 & 0.5579 & 0.6357 \\\hline
\end{tabular}
\end{table*}
\end{widetext}


\bibliography{eLSM_HSP}

\end{document}